\documentclass[aps,prl,twocolumn,longbibliography]{revtex4-1}

\usepackage{amsmath}
\usepackage{amssymb}
\usepackage{dsfont}
\usepackage{empheq}
\newcommand \unit{\mathds{1}}

\usepackage{xcolor}

\usepackage{setspace}
\usepackage{bm}
\usepackage{graphicx}
\usepackage{hyperref}
\usepackage[caption=false]{subfig}
\usepackage{adjustbox}
\usepackage{minitoc}

\AtBeginDocument{\let\oldcontentsline\contentsline}
\newcommand{\notoccontentsline}[4]{\oldcontentsline{}{}{}{}}
\newcommand{\droptocpage}{\addtocontents{toc}{\let\protect\contentsline\protect\notoccontentsline}}
\newcommand{\incltocpage}{\addtocontents{toc}{\let\protect\contentsline\protect\oldcontentsline}}

\begin{document}



\title{A Strongly Interacting Polaritonic Quantum Dot}
\author{Jia Ningyuan}
\author{Nathan Schine}
\author{Alexandros Georgakopoulos}
\author{Albert Ryou}
\author{Ariel Sommer}
\author{Jonathan Simon}
\affiliation{James Franck Institute and the Department of Physics at the University of Chicago}
\date{\today}

\begin{abstract}
Polaritons are an emerging platform for exploration of synthetic materials~\cite{carusotto2013quantum} and quantum information processing~\cite{Saffman_2010} that draw  properties from two disparate particles: a photon and an atom. Cavity polaritons are particularly promising, as they are long-lived and their dispersion and mass are controllable through cavity geometry~\cite{sommer2016engineering}. To date, studies of cavity polaritons have operated in the mean-field regime, using short-range interactions between their matter components~\cite{deng2010exciton}. Rydberg excitations have recently been demonstrated as a promising matter-component of polaritons~\cite{peyronel2012quantum}, due to their strong interactions over distances large compared to an optical wavelength. Here we explore, for the first time, the cavity quantum electrodynamics of Rydberg polaritons, combining the non-linearity of polaritonic quantum wires with the zero-dimensional strong coupling of an optical resonator. We assemble a quantum dot composed of $\sim 150$ strongly interacting, Rydberg-dressed $^{87}$Rb atoms in a cavity, and observe blockaded polariton transport as well as coherent quantum dynamics of a single polaritonic super-atom. This work establishes a new generation of photonic quantum information processors and quantum materials, along with a clear path to topological quantum matter\cite{schine2016synthetic}.
\end{abstract}

\maketitle

A strongly nonlinear resonator is the fundamental building block of both photonic quantum materials~\cite{carusotto2013quantum} and photonic quantum information processors~\cite{obrien2009photonic}. In the former case, coupling such resonators together yields an  interacting lattice model~\cite{hartmann2008quantum}, and more general resonator geometries give rise to photonic quantum Hall physics, frustration and glassy physics~\cite{gopalakrishnan2009emergent}, and long-range interactions~\cite{kollar2017supermode}. In the latter case, a single such resonator may act as either a quantum bit or quantum gate, with coupling between resonators providing light-speed information transport.

In recent years a number of different experimental platforms have emerged to realize single-photon-level non-linearities. In spite of their small absorption cross-section, individual atoms coupled to high-finesse, small-mode-volume optical resonators satisfy the requirements~\cite{Birnbaum2005,thompson2013coupling}, but face numerous engineering challenges in scaling up to multiple cavities or modes~\cite{goban2015superradiance}. Recently, it was demonstrated that atoms separated by micron-scale distances could be induced to interact through a Rydberg excited state~\cite{Urban2009,Gaetan2009}, enabling them to act as a ``super-atom'' with the absorption cross-section of many atoms and the non-linearity of a single emitter. In such a configuration, one-dimensional photonic quantum wires have been realized~\cite{peyronel2012quantum,dudin2012observation,dudi2012stro,tiar2014sing,gorn2014sing}, where individual photons collide with high probability.

An appealing possibility is to marry the two approaches~\cite{Guerlin2010}, employing Rydberg ``super-atoms'' in an optical resonator. Prior efforts in this direction~\cite{pari2012obse,ningyuan2016observation} have achieved weak mean-field interactions of many photons; here, for the first time, we enter the regime of strong interactions between individual photons by harnessing advances in resonator design (see SI \ref{SI:CavityDetails}) that mitigate coupling of the delicate Rydberg atoms to nearby surfaces. We realize a \emph{zero}-dimensional quantum dot, creating a versatile strongly interacting platform for quantum information processing and materials synthesis~\cite{sommer2016engineering,anderson2016engineering,gorshkov2011photon}. 

In the weakly interacting regime, resonators have already been employed to explore a number of phenomena, including non-local and frustrated interactions between atoms ~\cite{gopalakrishnan2009emergent,baumann2010dicke,leonard2016supersolid}, quantum-degenerate fluids~\cite{deng2010exciton,klaers2010bose}, and Landau levels on curved manifolds~\cite{schine2016synthetic}.  The ease of injecting and selectively removing photons has enabled a new generation of dissipatively-engineered materials~\cite{raftery2014observation}, along with proposals to stabilize more exotic phases~\cite{hafezi2015,lebreuilly2016towards,ma2017autonomous}.

In what follows, we describe our cavity Rydberg polariton-based quantum dot and show that it exhibits the defining features of strong nonlinearity at the single-photon level. We begin by demonstrating strong light-matter coupling via spectrally isolated Rydberg-polariton resonances and then probe the strong interactions between individual polaritons through transport blockade, single-polariton Rabi oscillations and ring-down of the dot's occupation. We conclude with a discussion of applications in quantum information, as well as strongly-correlated and topological phases of matter, explored through the use of recently developed dissipative engineering tools.


\begin{figure*}[t]
	\includegraphics[width=2\columnwidth]{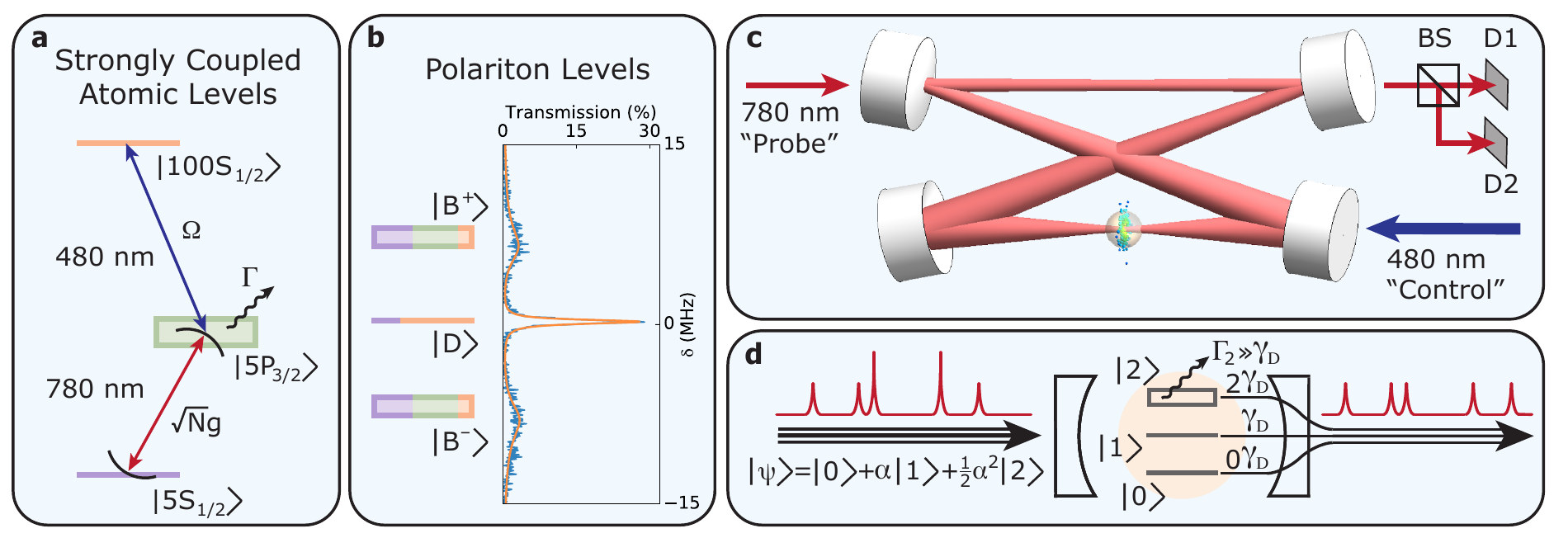}
	\caption{\textbf{Rydberg Polariton Blockade in an Optical Resonator.} \textbf{(a)} Because photons do not naturally interact with one another, we employ atoms to mediate interactions. We employ an optical resonator to couple 780 nm photons to a sample of $N \approx$ 150 $^{87}$Rb atoms, thereby benefiting from a collective $\sqrt{N}g$ enhancement in the single atom-photon Rabi frequency, $g$. After a probe photon is absorbed collectively by the atomic ensemble, the atoms are driven coherently to the $n$ = 100 Rydberg state using a 480 nm ``control'' laser with Rabi frequency $\Omega$. Rydberg atoms interact strongly with one another, and the light-matter coupling imprints these interactions on the $780$nm photons. \textbf{(b)} When we probe the spectrum of this strongly-coupled light-matter system, we observe three distinct peaks corresponding to the three quasi-particle eigenstates, called polaritons: two broad ``bright'' polaritons, with a linewidth set by the excited state spontaneous decay rate $\Gamma$, and one narrow, tall “dark” polariton in the middle, with a loss rate $\gamma_{D}$ independent of $\Gamma$. The dark polariton is the narrowest and most Rydberg-like quasi-particle, so it provides the best platform for mediating interactions between photons. \textbf{(c)} Atoms from a magneto-optical trap (not shown) are loaded into the smallest waist of an optical resonator, which is itself resonant with the 780 nm light. Each photon that enters the resonator hybridizes with the atomic sample to create a dark Rydberg polariton that excludes any other Rydberg excitations within a surrounding ``blockade volume''. To ensure that any pair of photons interacts strongly, it is necessary to spatially restrict the atoms to a volume less than this blockade volume. The atomic cloud is “sliced” longitudinally (along the resonator axis) by spatially selective optical depumping (see SI \ref{SI:AtomSlicing}) so that, in combination with the radial selectivity of the resonator mode, all photons in the resonator create a Rydberg excitation within a single blockade volume and therefore interact strongly with one another. Signatures of this interaction are then observed in the 780 nm photons that leak out through one of the resonator mirrors, pass through a 50:50 beam-splitter (BS) and are measured by single photon detectors (D1 and D2). \textbf{(d)}. The system behaves as a zero-dimensional strongly interacting quantum dot. The weak probe field is in a coherent state of some small parameter $\alpha$, represented by the red curve which consists primarily of zero photons, occasionally one photon, and infrequently two photons, when counts are binned into a time $1/\gamma_{D}$. Each photon enters the cavity as a dark polariton, and strong Rydberg-Rydberg interactions cause the first polariton to broaden and shift the dark polariton resonance sufficiently to preclude injection of a second polariton simultaneously; accordingly, only isolated photons emerge from the cavity.\label{Figure:SetupFigure}
}
\end{figure*}

\begin{figure}[t]
	\centering
	\includegraphics[width=0.9\columnwidth]{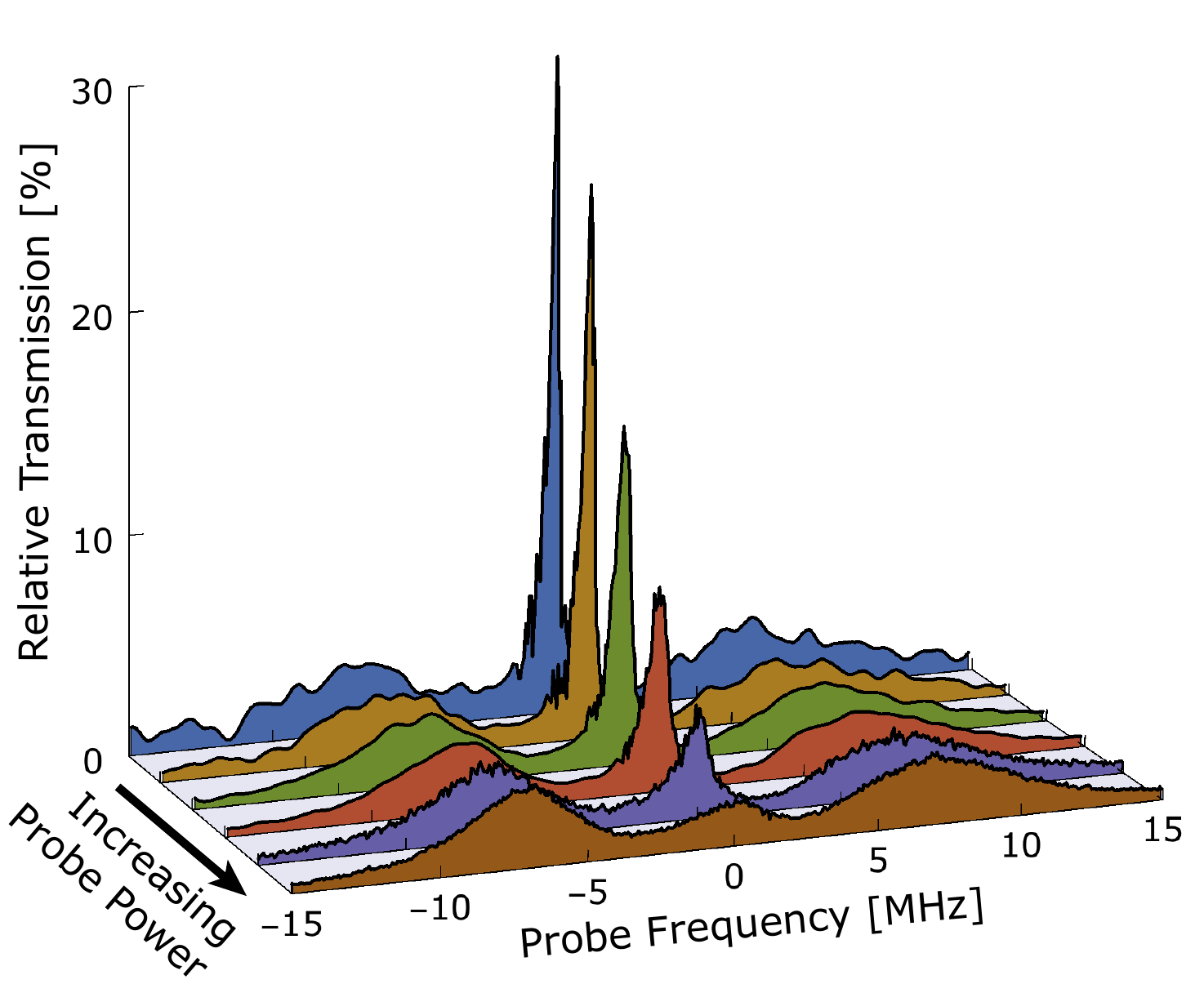}
	\caption{\textbf{Nonlinear Spectroscopy of a Polaritonic Quantum Dot.} The quantum dot's transmission versus probe frequency is plotted for varying incident photon rate from 320 ms$^{-1}$ to 98.8 $\mu$s$^{-1}$ in log-uniform steps. Because the atomic sample employed in the present dot is unsliced and thus comprises several ($\sim 3$) blockade volumes, interactions are nearly mean-field; we thus observe smooth broadening and transmission suppression of the dark polariton peak (at $\delta=0$ MHz). The bright polariton resonances at $\delta=\pm 7$ MHz are not suppressed at high probe rate, because bright polaritons have little Rydberg admixture and are thus essentially non-interacting. The data for the three lowest powers are Gaussian filtered (with $\sigma_{filter}=481$ kHz) outside of the interval (-1.5 MHz, 1.5 MHz); all data are normalized to the bare cavity transmission.\label{Fig:EITvsPrbPwr}}
\end{figure}

\begin{figure}[h!]
	\subfloat{
		\includegraphics{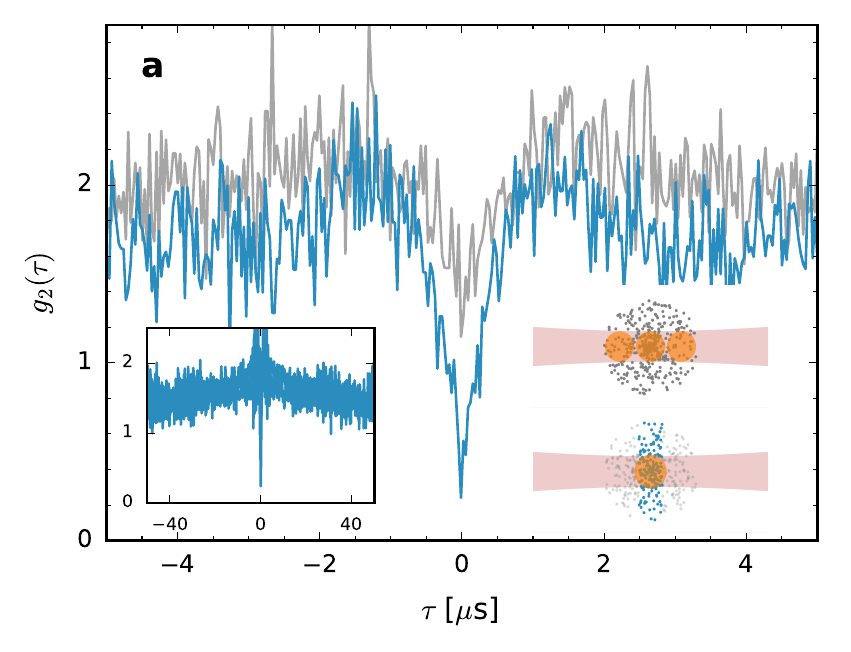}
	}\\[-4ex]
	\subfloat{
		\includegraphics{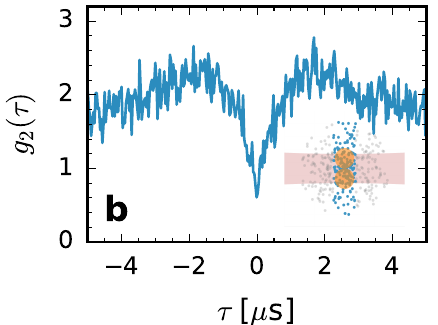}
    }%
    \subfloat{
		\includegraphics{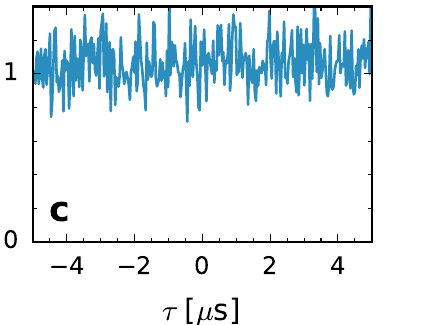}
    }
    \label{Figure:g2}
	\caption{\label{Figure:g2} \textbf{Transport Blockade of Cavity Rydberg Polaritons.} \textbf{(a)} A smoking gun of strong interactions in a quantum dot is suppression of near-simultaneous particle-tunneling through it. In our polaritonic dot, this manifests as a suppression of near-simultaneous transit of photon pairs through the resonator, quantified by the temporal intensity autocorrelation function $g_2(\tau)$; this function compares the rate of photons leaking out of the resonator separated by a time $\tau$ to a Poisson-process with the same average rate. While probing the cavity on the peak of the dark polariton resonance in a sliced, 10 $\mu$m RMS length atomic cloud, we observe a $\sim 5$x suppression of two-photon events relative to the long-time value depicted in the inset (blue curve, $g_2=0.27(8)$ near $\tau=0$ compared to $g_2=1.3$ at long times). The unsliced cloud length is $37$ $\mu$m RMS, containing numerous blockade volumes; accordingly it provides a much weaker suppression of $g_2$ near $\tau=0$ (gray). That it provides any suppression at all is only due to the mutual coupling of all blockade volumes to the same resonator mode. The common $900$ ns exponential rise time of the $g_2$ indicates equal dark polariton linewidths in the sliced and unsliced clouds. \textbf{(b)} $g_2(\tau)$ is plotted on the dark polariton resonance of the 85S$_{1/2}$ Rydberg state with a sliced cloud, where the reduced blockade radius begins to allow multiple blockade volumes in the transverse plane. \textbf{(c)} $g_2(\tau)$ is plotted while probing a bright polariton resonance at $n=100$. Decreased Rydberg state proportion reduces the interaction strength while the significant excited state contribution substantially increases the polariton linewidth;  the excitation spectrum thereby becomes linear and no sub-Poissonian statistics are observed. For \textbf{a} and \textbf{b}, the large-$\tau$ value of $g_2$ exceeds 1 due to atom number fluctuations, which lead to a variation in the on-resonance transmission. In the inset to \textbf{a}, the time-bins for $|\tau|>0.8$ $\mu$s are larger by a factor 4 to reveal large $\tau$ trends.}
\end{figure}

\begin{figure*}[t]
\subfloat{
	\includegraphics{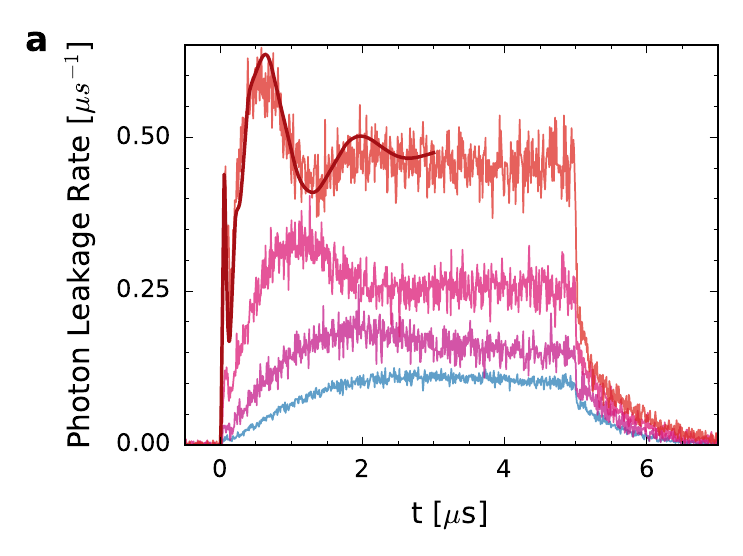}
}\hspace{-3ex}
\subfloat{
	\includegraphics{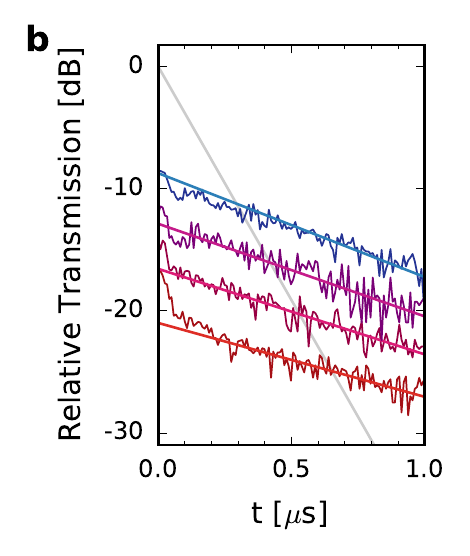}
}\hspace{-5ex}
\subfloat{
	\includegraphics{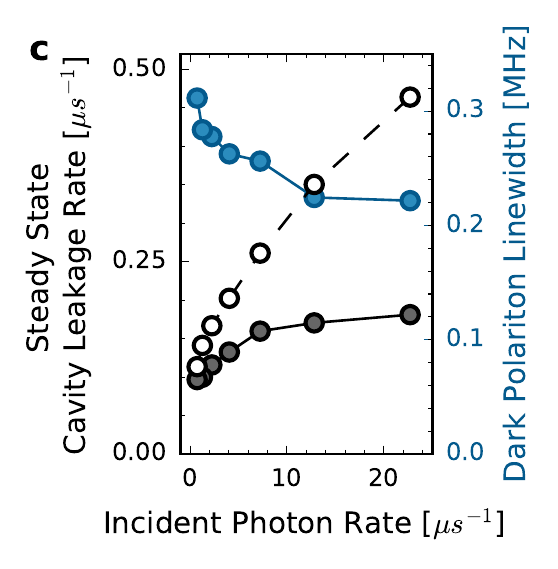}
}
\caption{\label{Figure:ringdown} \textbf{Dynamical Evolution of a Polaritonic Quantum Dot: Ring-up and Ring-down.} When the dot is simultaneously excited with many photons within the dark polariton lifetime, these photons stimulate coherent tunneling of a polariton into and out-of the dot, before the dot finally equilibrates. In \textbf{(a)}, we excite the resonator on the dark polariton resonance for 5 $\mu$s and record the transmitted intensity during the equilibration process, and subsequent ring-down. As the probe power is increased (blue to red), the overall transmitted power increases and a significant oscillation develops, reflecting the Rabi oscillation of the dot between zero and one polariton states. The ring-up of the resonator is compared to master equation numerics (see SI ~\ref{SI:SimpleMasterEqn}). The slow dynamics ($\sim\mu$s timescale) reflect the tunneling oscillations between zero and one dark polaritons, while the fast dynamics (on a $\sim 100$ ns timescale) reflect weak off-resonant excitation of the broad bright polaritons. Bright polaritons are very light-like (see Fig. \ref{Figure:SetupFigure}), and thus interact only weakly and do not saturate with increasing probe power. Accordingly, they produce a growing background at the highest probe powers (nearly 50\% of the observed signal), which weakens the observed saturation of the resonator transmission (see panel \textbf{c}). \textbf{(b)} To isolate the physics of the dark polariton saturation from the bright polariton background, we explore the ring-down of the dot once the probe field removed. The first $\sim 100$ ns of ringdown is dominated by the fast decay of bright polaritons, with oscillations arising from  interference of upper and lower bright polaritons. At later times, the cavity leakage is dominated entirely by the dark polaritons, with a lifetime substantially longer than the bare resonator (gray). To learn about the steady-state dark polariton number, we extrapolate the slow decay of cavity emission to the beginning of the ring-down. \textbf{(c)} The \emph{actual} cavity-emission at the beginning of the ring-down reflects both bright and dark polariton occupation in the dot, and does not saturate with increasing probe power (white, open); by contrast, the extrapolated zero-time dark polariton emission rate saturates strongly (black, solid) and its linewidth decreases (blue). The saturated dark polariton \emph{number} at the highest probe power is $n_{D}=0.16(8)$  (see SI \ref{SI:DSPnumber}).}
\end{figure*}

We load a sample of $2600(500)$ atoms into the $12\mu$m$\times 14\mu$m waist of a high finesse single mode optical resonator (Fig. \ref{Figure:SetupFigure}a,c), at a peak density of $1\times 10^{11}$cm$^{-3}$. These atoms are distributed over a 35 $\mu$m RMS axial length which may be reduced to 10 $\mu$m RMS by spatially selective optical depumping, or ``slicing'' (see SI ~\ref{SI:AtomSlicing}). Due to strong light-matter coupling, the modes of the system hybridize, forming polaritons: composite states of a resonator photon and an atomic excitation. One of the polaritons, called the ``dark'' polariton, consists primarily of a Rydberg excitation admixed weakly with a resonator photon (Fig. \ref{Figure:SetupFigure}b). The strong repulsion between nearby Rydberg atoms enables dark polaritons to interact strongly with one another. The tightly confined geometry of the resonator mode and atomic sample gives rise to a zero-dimensional, strongly interacting polaritonic quantum dot; like its solid-state counterpart, the electronic quantum dot~\cite{kouwenhoven1997electron}, it exhibits blockaded transport (Fig. \ref{Figure:SetupFigure}d).

To investigate the properties of our polaritonic quantum dot, we first probe the excitation spectrum of the laser-dressed cavity-atom system: in Fig. \ref{Figure:SetupFigure}b we plot the quantum dot's transmission spectrum on the $\{5S_{1/2} F=2\}\rightarrow \{5P_{3/2}F=3\}\rightarrow \{100S_{1/2}\}$ transition, without slicing the cloud. We observe two spectrally-broad ``bright'' polaritons composed primarily of an admixture of a resonator photon and a P-state atom, whose spectral separation is an indication of strong light-matter coupling. The central feature, however, is the spectrally narrow dark polariton, composed of a coherent superposition of a $100S$ excitation, and a resonator photon, with the mixing ratio set by the dark-state rotation angle~\cite{fleischhauer2005electromagnetically}: $\tan{\theta}=\sqrt{N}g/\Omega=2.9(5)$, for our typical conditions. The Rydberg component of the dark polariton enables it to strongly repel other dark polaritons within a surrounding ``blockade volume,'' which scales with $n^{\frac{11}{2}}$ where $n$ is the Rydberg level's principal quantum number. It is for this reason that we employ $n=100$ Rydberg atoms, as their extreme properties make this blockade radius on the order of the sample size. Indeed, as we increase the probe power, the fractional transmission on the dark polariton resonance drops, as shown in Fig. ~\ref{Fig:EITvsPrbPwr}; this indicates that the strong repulsion between polaritons suppresses their simultaneous injection into the dot.

In order to demonstrate that the dark polaritons interact strongly with one another within the quantum dot, we explore their simultaneous transport through it. We achieve this by injecting photons into the resonator at the energy of the dark-polariton feature, such that the photons become dark polaritons upon entering. The strong repulsion between polaritons shifts the energy and reduces the lifetime of a second polariton in the dot, thereby precluding its injection. In an electronic quantum dot the analogous Coulomb-blockade physics is typically ascertained from the dependence of transport upon bias-voltage~\cite{kouwenhoven1997electron}; by contrast, we directly observe suppression of simultaneous polariton transit by detecting \emph{when} photons tunnel through the dot. This is achieved via the temporal intensity autocorrelation function, or $g_2$, of the cavity transmission, plotted in Fig. \ref{Figure:g2}: a suppression near $\tau=0$ indicates interaction-driven suppression of double-occupancy of the dot, while the rise-time back to steady-state reflects the smaller of the drive Rabi-frequency and polariton linewidth. In Fig. \ref{Figure:g2}a the unsliced cloud (gray trace) exhibits only weak suppression of $g_2$ near $\tau=0$ because this dot is large enough to hold multiple polaritons along the resonator axis simultaneously; the sliced cloud (blue trace) exhibits a strong suppression ($g_2=0.27(8)$) because it can hold only a single polariton at a time. In both cases, the $\sim\mu$s-timescale corresponds to a polariton linewidth slightly narrower than the $2\pi\times 400$kHz observed in Fig. \ref{Figure:SetupFigure}b, indicative of slight inhomogeneous broadening (see SI ~\ref{SI:inhomog}). In Fig. \ref{Figure:g2}b, we explore the $g_2$ of a polaritonic dot in the $85S$ Rydberg state, which exhibits yet weaker $g_2$ suppression because this dot can support multiple excitations in the plane transverse to the resonator axis. To further demonstrate that we are exploring the physics of polaritons composed of \emph{many} strongly interacting atoms, and not single-atom Jaynes-Cummings physics~\cite{Birnbaum2005}, in Fig. \ref{Figure:g2}c we plot $g_2(\tau)$ while probing on a bright polariton resonance, and observe no suppression of the $g_2$ whatsoever. Bright polaritons are weakly interacting due to their reduced Rydberg component and increased linewidth, so no anti-bunching is expected here, except in the case of precisely one atom within the resonator.


To investigate the coherent dynamics of the polaritonic dot, we probe it with a laser pulse at the energy of the 100S dark polariton resonance and observe the transmitted light as the intra-cavity polariton field rings up and then down. Figure \ref{Figure:ringdown}a shows the ring-up dynamics for various probe intensities, exhibiting Rabi oscillations between zero and one dark polaritons at the highest intensities, indicative of a strongly blockaded dot interacting with many photons within the polariton lifetime. The dark polariton oscillations exhibit a Rabi frequency of $\Omega_{polariton}=2\pi\times 318$ kHz, in agreement with a first-principles calculation based upon the probe power (see SI ~\ref{SI:DSPrabifreq}). The solid curve is the numerical integration of a simple master-equation model (see SI ~\ref{SI:SimpleMasterEqn}) allowing up to two interacting Rydberg excitations within the system. The fast $\sim 100$ns Rabi oscillations arise from off-resonant excitation of non-interacting bright polaritons, and as such their waveform is simply proportional to probe power, growing in proportion to the dark polariton signal which saturates due to the strong interactions between dark polaritons. To disentangle the bright and dark contributions to the signal, Fig. ~\ref{Figure:ringdown}b shows the ring-down of the dot once the probe beam is turned off. At each probe intensity, the curve consists of slow and fast exponential decays; the slow feature (with time constant of $\tau=590(70)$ns), reflects the dynamics of the dark polariton, and the fast dynamics reflect decay of the two interfering bright polariton modes. Figure \ref{Figure:ringdown}c shows the extrapolation of the slow decay of the cavity emission to zero-time (solid, black curve), plotted against the drive power. The strong saturation with drive power again indicates the blockade of the dark polariton manifold. To further demonstrate this saturation, we measure the $g_2$ of the slowly-decaying tail of the ring-down (after the bright polaritons have all decayed away), and find $g_2=0.31(7)$.


In this work we have demonstrated, for the first time, strong interactions between individual cavity Rydberg polaritons. Beginning with a spectroscopic demonstration of well-resolved polaritonic quasi-particles in the resonator transmission spectrum, we explore the interactions between dark polaritons via the statistics of photons tunneling through the cavity; strong anti-bunching at zero-time separation validates a model where a single intra-cavity polariton shifts and broadens the energy for the injection of the next polariton by more than the polariton linewidth, strongly suppressing its tunneling into the resonator until the first polariton tunnels out. We are further able to observe coherent tunneling of a single polariton into and out of the resonator when the system is simultaneously subjected to many photons within the dark polariton lifetime.

Cavity polaritons are now ripe for applications in both quantum information processing~\cite{zhao2010efficient,brion2012quantum} and synthetic quantum materials. Recent work has demonstrated synthetic magnetic fields for resonator photons ~\cite{schine2016synthetic}; in conjunction with the present work, there is now a clear path to strongly correlated photonic quantum materials ~\cite{carusotto2013quantum,bienias2016few}. Broadly, the numerous proposals to study synthetic quantum matter in cavity arrays and continua~\cite{carusotto2013quantum} can now be explored in the modes of a single multi-mode optical cavity, employing Rydberg atoms to mediate strong photon-photon interactions; achieving stronger light-matter coupling will require only additional atomic density, rather than increasingly challenging advances in optical super-mirror technology necessary for single-emitter approaches. Upcoming challenges will center upon harnessing these developments to answer questions in dissipative preparation of manybody quantum states ~\cite{hafezi2015,lebreuilly2016towards,ma2017autonomous}, and exploration of the resulting phase diagrams~\cite{grusdt2013fractional} with manipulation and detection~\cite{grusdt2016interferometric,umucalilar2013many} unique to quantum optics.

\droptocpage
\section{Acknowledgements}
\label{sec:Acknowledgements}
We would like to thank Michael Fleischhauer and Hanspeter Buechler for fruitful conversations. This work was supported by DOE grant DE-SC0010267 for apparatus construction, DARPA grant W911NF-15-1-0620 for modeling, and MURI grant FA9550-16-1-0323 for data collection and analysis. A.G. acknowledges support from the UChicago MRSEC Grant NSF-DMR-MRSEC 1420709. A.R. acknowledges support from the NDSEG Fellowship.

\section{Author Contributions}
The experiment was designed and built by all authors. J.N., N.S., and J.S. collected and analyzed the data. All authors contributed to the manuscript.

\section{Author Information}
The authors declare no competing financial interests. Correspondence and requests for materials should be addressed to J.S. (simonjon@uchicago.edu)

\bibliography{BibliographyStuff/TheBib}{}

\begin{thebibliography}{41}%
\makeatletter
\providecommand \@ifxundefined [1]{%
 \@ifx{#1\undefined}
}%
\providecommand \@ifnum [1]{%
 \ifnum #1\expandafter \@firstoftwo
 \else \expandafter \@secondoftwo
 \fi
}%
\providecommand \@ifx [1]{%
 \ifx #1\expandafter \@firstoftwo
 \else \expandafter \@secondoftwo
 \fi
}%
\providecommand \natexlab [1]{#1}%
\providecommand \enquote  [1]{``#1''}%
\providecommand \bibnamefont  [1]{#1}%
\providecommand \bibfnamefont [1]{#1}%
\providecommand \citenamefont [1]{#1}%
\providecommand \href@noop [0]{\@secondoftwo}%
\providecommand \href [0]{\begingroup \@sanitize@url \@href}%
\providecommand \@href[1]{\@@startlink{#1}\@@href}%
\providecommand \@@href[1]{\endgroup#1\@@endlink}%
\providecommand \@sanitize@url [0]{\catcode `\\12\catcode `\$12\catcode
  `\&12\catcode `\#12\catcode `\^12\catcode `\_12\catcode `\%12\relax}%
\providecommand \@@startlink[1]{}%
\providecommand \@@endlink[0]{}%
\providecommand \url  [0]{\begingroup\@sanitize@url \@url }%
\providecommand \@url [1]{\endgroup\@href {#1}{\urlprefix }}%
\providecommand \urlprefix  [0]{URL }%
\providecommand \Eprint [0]{\href }%
\providecommand \doibase [0]{http://dx.doi.org/}%
\providecommand \selectlanguage [0]{\@gobble}%
\providecommand \bibinfo  [0]{\@secondoftwo}%
\providecommand \bibfield  [0]{\@secondoftwo}%
\providecommand \translation [1]{[#1]}%
\providecommand \BibitemOpen [0]{}%
\providecommand \bibitemStop [0]{}%
\providecommand \bibitemNoStop [0]{.\EOS\space}%
\providecommand \EOS [0]{\spacefactor3000\relax}%
\providecommand \BibitemShut  [1]{\csname bibitem#1\endcsname}%
\let\auto@bib@innerbib\@empty
\bibitem [{\citenamefont {Carusotto}\ and\ \citenamefont
  {Ciuti}(2013)}]{carusotto2013quantum}%
  \BibitemOpen
  \bibfield  {author} {\bibinfo {author} {\bibfnamefont {Iacopo}\ \bibnamefont
  {Carusotto}}\ and\ \bibinfo {author} {\bibfnamefont {Cristiano}\ \bibnamefont
  {Ciuti}},\ }\bibfield  {title} {\enquote {\bibinfo {title} {Quantum fluids of
  light},}\ }\href@noop {} {\bibfield  {journal} {\bibinfo  {journal} {Reviews
  of Modern Physics}\ }\textbf {\bibinfo {volume} {85}},\ \bibinfo {pages}
  {299} (\bibinfo {year} {2013})}\BibitemShut {NoStop}%
\bibitem [{\citenamefont {Saffman}\ \emph {et~al.}(2010)\citenamefont
  {Saffman}, \citenamefont {Walker},\ and\ \citenamefont
  {M\o{}lmer}}]{Saffman_2010}%
  \BibitemOpen
  \bibfield  {author} {\bibinfo {author} {\bibfnamefont {M.}~\bibnamefont
  {Saffman}}, \bibinfo {author} {\bibfnamefont {T.~G.}\ \bibnamefont {Walker}},
  \ and\ \bibinfo {author} {\bibfnamefont {K.}~\bibnamefont {M\o{}lmer}},\
  }\bibfield  {title} {\enquote {\bibinfo {title} {Quantum information with
  rydberg atoms},}\ }\href {\doibase 10.1103/RevModPhys.82.2313} {\bibfield
  {journal} {\bibinfo  {journal} {Rev. Mod. Phys.}\ }\textbf {\bibinfo {volume}
  {82}},\ \bibinfo {pages} {2313--2363} (\bibinfo {year} {2010})}\BibitemShut
  {NoStop}%
\bibitem [{\citenamefont {Sommer}\ and\ \citenamefont
  {Simon}(2016)}]{sommer2016engineering}%
  \BibitemOpen
  \bibfield  {author} {\bibinfo {author} {\bibfnamefont {Ariel}\ \bibnamefont
  {Sommer}}\ and\ \bibinfo {author} {\bibfnamefont {Jonathan}\ \bibnamefont
  {Simon}},\ }\bibfield  {title} {\enquote {\bibinfo {title} {Engineering
  photonic floquet hamiltonians through fabry--p{\'e}rot resonators},}\
  }\href@noop {} {\bibfield  {journal} {\bibinfo  {journal} {New Journal of
  Physics}\ }\textbf {\bibinfo {volume} {18}},\ \bibinfo {pages} {035008}
  (\bibinfo {year} {2016})}\BibitemShut {NoStop}%
\bibitem [{\citenamefont {Deng}\ \emph {et~al.}(2010)\citenamefont {Deng},
  \citenamefont {Haug},\ and\ \citenamefont {Yamamoto}}]{deng2010exciton}%
  \BibitemOpen
  \bibfield  {author} {\bibinfo {author} {\bibfnamefont {Hui}\ \bibnamefont
  {Deng}}, \bibinfo {author} {\bibfnamefont {Hartmut}\ \bibnamefont {Haug}}, \
  and\ \bibinfo {author} {\bibfnamefont {Yoshihisa}\ \bibnamefont {Yamamoto}},\
  }\bibfield  {title} {\enquote {\bibinfo {title} {Exciton-polariton
  bose-einstein condensation},}\ }\href@noop {} {\bibfield  {journal} {\bibinfo
   {journal} {Reviews of Modern Physics}\ }\textbf {\bibinfo {volume} {82}},\
  \bibinfo {pages} {1489} (\bibinfo {year} {2010})}\BibitemShut {NoStop}%
\bibitem [{\citenamefont {Peyronel}\ \emph {et~al.}(2012)\citenamefont
  {Peyronel}, \citenamefont {Firstenberg}, \citenamefont {Liang}, \citenamefont
  {Hofferberth}, \citenamefont {Gorshkov}, \citenamefont {Pohl}, \citenamefont
  {Lukin},\ and\ \citenamefont {Vuleti{\'c}}}]{peyronel2012quantum}%
  \BibitemOpen
  \bibfield  {author} {\bibinfo {author} {\bibfnamefont {Thibault}\
  \bibnamefont {Peyronel}}, \bibinfo {author} {\bibfnamefont {Ofer}\
  \bibnamefont {Firstenberg}}, \bibinfo {author} {\bibfnamefont {Qi-Yu}\
  \bibnamefont {Liang}}, \bibinfo {author} {\bibfnamefont {Sebastian}\
  \bibnamefont {Hofferberth}}, \bibinfo {author} {\bibfnamefont {Alexey~V}\
  \bibnamefont {Gorshkov}}, \bibinfo {author} {\bibfnamefont {Thomas}\
  \bibnamefont {Pohl}}, \bibinfo {author} {\bibfnamefont {Mikhail~D}\
  \bibnamefont {Lukin}}, \ and\ \bibinfo {author} {\bibfnamefont {Vladan}\
  \bibnamefont {Vuleti{\'c}}},\ }\bibfield  {title} {\enquote {\bibinfo {title}
  {Quantum nonlinear optics with single photons enabled by strongly interacting
  atoms},}\ }\href@noop {} {\bibfield  {journal} {\bibinfo  {journal} {Nature}\
  }\textbf {\bibinfo {volume} {488}},\ \bibinfo {pages} {57--60} (\bibinfo
  {year} {2012})}\BibitemShut {NoStop}%
\bibitem [{\citenamefont {Schine}\ \emph {et~al.}(2016)\citenamefont {Schine},
  \citenamefont {Ryou}, \citenamefont {Gromov}, \citenamefont {Sommer},\ and\
  \citenamefont {Simon}}]{schine2016synthetic}%
  \BibitemOpen
  \bibfield  {author} {\bibinfo {author} {\bibfnamefont {Nathan}\ \bibnamefont
  {Schine}}, \bibinfo {author} {\bibfnamefont {Albert}\ \bibnamefont {Ryou}},
  \bibinfo {author} {\bibfnamefont {Andrey}\ \bibnamefont {Gromov}}, \bibinfo
  {author} {\bibfnamefont {Ariel}\ \bibnamefont {Sommer}}, \ and\ \bibinfo
  {author} {\bibfnamefont {Jonathan}\ \bibnamefont {Simon}},\ }\bibfield
  {title} {\enquote {\bibinfo {title} {Synthetic landau levels for photons},}\
  }\href@noop {} {\bibfield  {journal} {\bibinfo  {journal} {Nature}\ }\textbf
  {\bibinfo {volume} {534}},\ \bibinfo {pages} {671} (\bibinfo {year}
  {2016})}\BibitemShut {NoStop}%
\bibitem [{\citenamefont {O'brien}\ \emph {et~al.}(2009)\citenamefont
  {O'brien}, \citenamefont {Furusawa},\ and\ \citenamefont
  {Vu{\v{c}}kovi{\'c}}}]{obrien2009photonic}%
  \BibitemOpen
  \bibfield  {author} {\bibinfo {author} {\bibfnamefont {Jeremy~L}\
  \bibnamefont {O'brien}}, \bibinfo {author} {\bibfnamefont {Akira}\
  \bibnamefont {Furusawa}}, \ and\ \bibinfo {author} {\bibfnamefont {Jelena}\
  \bibnamefont {Vu{\v{c}}kovi{\'c}}},\ }\bibfield  {title} {\enquote {\bibinfo
  {title} {Photonic quantum technologies},}\ }\href@noop {} {\bibfield
  {journal} {\bibinfo  {journal} {Nature Photonics}\ }\textbf {\bibinfo
  {volume} {3}},\ \bibinfo {pages} {687--695} (\bibinfo {year}
  {2009})}\BibitemShut {NoStop}%
\bibitem [{\citenamefont {Hartmann}\ \emph {et~al.}(2008)\citenamefont
  {Hartmann}, \citenamefont {Brandao},\ and\ \citenamefont
  {Plenio}}]{hartmann2008quantum}%
  \BibitemOpen
  \bibfield  {author} {\bibinfo {author} {\bibfnamefont {Michael~J}\
  \bibnamefont {Hartmann}}, \bibinfo {author} {\bibfnamefont {Fernando~GSL}\
  \bibnamefont {Brandao}}, \ and\ \bibinfo {author} {\bibfnamefont {Martin~B}\
  \bibnamefont {Plenio}},\ }\bibfield  {title} {\enquote {\bibinfo {title}
  {Quantum many-body phenomena in coupled cavity arrays},}\ }\href@noop {}
  {\bibfield  {journal} {\bibinfo  {journal} {Laser \& Photonics Reviews}\
  }\textbf {\bibinfo {volume} {2}},\ \bibinfo {pages} {527--556} (\bibinfo
  {year} {2008})}\BibitemShut {NoStop}%
\bibitem [{\citenamefont {Gopalakrishnan}\ \emph {et~al.}(2009)\citenamefont
  {Gopalakrishnan}, \citenamefont {Lev},\ and\ \citenamefont
  {Goldbart}}]{gopalakrishnan2009emergent}%
  \BibitemOpen
  \bibfield  {author} {\bibinfo {author} {\bibfnamefont {Sarang}\ \bibnamefont
  {Gopalakrishnan}}, \bibinfo {author} {\bibfnamefont {Benjamin~L}\
  \bibnamefont {Lev}}, \ and\ \bibinfo {author} {\bibfnamefont {Paul~M}\
  \bibnamefont {Goldbart}},\ }\bibfield  {title} {\enquote {\bibinfo {title}
  {Emergent crystallinity and frustration with bose--einstein condensates in
  multimode cavities},}\ }\href@noop {} {\bibfield  {journal} {\bibinfo
  {journal} {Nature Physics}\ }\textbf {\bibinfo {volume} {5}},\ \bibinfo
  {pages} {845--850} (\bibinfo {year} {2009})}\BibitemShut {NoStop}%
\bibitem [{\citenamefont {Koll{\'a}r}\ \emph {et~al.}(2017)\citenamefont
  {Koll{\'a}r}, \citenamefont {Papageorge}, \citenamefont {Vaidya},
  \citenamefont {Guo}, \citenamefont {Keeling},\ and\ \citenamefont
  {Lev}}]{kollar2017supermode}%
  \BibitemOpen
  \bibfield  {author} {\bibinfo {author} {\bibfnamefont {Alicia~J}\
  \bibnamefont {Koll{\'a}r}}, \bibinfo {author} {\bibfnamefont {Alexander~T}\
  \bibnamefont {Papageorge}}, \bibinfo {author} {\bibfnamefont {Varun~D}\
  \bibnamefont {Vaidya}}, \bibinfo {author} {\bibfnamefont {Yudan}\
  \bibnamefont {Guo}}, \bibinfo {author} {\bibfnamefont {Jonathan}\
  \bibnamefont {Keeling}}, \ and\ \bibinfo {author} {\bibfnamefont
  {Benjamin~L}\ \bibnamefont {Lev}},\ }\bibfield  {title} {\enquote {\bibinfo
  {title} {Supermode-density-wave-polariton condensation with a bose--einstein
  condensate in a multimode cavity},}\ }\href@noop {} {\bibfield  {journal}
  {\bibinfo  {journal} {Nature Communications}\ }\textbf {\bibinfo {volume}
  {8}} (\bibinfo {year} {2017})}\BibitemShut {NoStop}%
\bibitem [{\citenamefont {Birnbaum}\ \emph {et~al.}(2005)\citenamefont
  {Birnbaum}, \citenamefont {Boca}, \citenamefont {Miller}, \citenamefont
  {Boozer}, \citenamefont {Northup},\ and\ \citenamefont
  {Kimble}}]{Birnbaum2005}%
  \BibitemOpen
  \bibfield  {author} {\bibinfo {author} {\bibfnamefont {K.~M.}\ \bibnamefont
  {Birnbaum}}, \bibinfo {author} {\bibfnamefont {A.}~\bibnamefont {Boca}},
  \bibinfo {author} {\bibfnamefont {R.}~\bibnamefont {Miller}}, \bibinfo
  {author} {\bibfnamefont {A.~D.}\ \bibnamefont {Boozer}}, \bibinfo {author}
  {\bibfnamefont {T.~E.}\ \bibnamefont {Northup}}, \ and\ \bibinfo {author}
  {\bibfnamefont {H.~J.}\ \bibnamefont {Kimble}},\ }\bibfield  {title}
  {\enquote {\bibinfo {title} {Photon blockade in an optical cavity with one
  trapped atom},}\ }\href {\doibase 10.1038/nature03804} {\bibfield  {journal}
  {\bibinfo  {journal} {Nature}\ }\textbf {\bibinfo {volume} {436}},\ \bibinfo
  {pages} {87--90} (\bibinfo {year} {2005})}\BibitemShut {NoStop}%
\bibitem [{\citenamefont {Thompson}\ \emph {et~al.}(2013)\citenamefont
  {Thompson}, \citenamefont {Tiecke}, \citenamefont {de~Leon}, \citenamefont
  {Feist}, \citenamefont {Akimov}, \citenamefont {Gullans}, \citenamefont
  {Zibrov}, \citenamefont {Vuleti{\'c}},\ and\ \citenamefont
  {Lukin}}]{thompson2013coupling}%
  \BibitemOpen
  \bibfield  {author} {\bibinfo {author} {\bibfnamefont {JD}~\bibnamefont
  {Thompson}}, \bibinfo {author} {\bibfnamefont {TG}~\bibnamefont {Tiecke}},
  \bibinfo {author} {\bibfnamefont {NP}~\bibnamefont {de~Leon}}, \bibinfo
  {author} {\bibfnamefont {J}~\bibnamefont {Feist}}, \bibinfo {author}
  {\bibfnamefont {AV}~\bibnamefont {Akimov}}, \bibinfo {author} {\bibfnamefont
  {M}~\bibnamefont {Gullans}}, \bibinfo {author} {\bibfnamefont
  {AS}~\bibnamefont {Zibrov}}, \bibinfo {author} {\bibfnamefont
  {V}~\bibnamefont {Vuleti{\'c}}}, \ and\ \bibinfo {author} {\bibfnamefont
  {MD}~\bibnamefont {Lukin}},\ }\bibfield  {title} {\enquote {\bibinfo {title}
  {Coupling a single trapped atom to a nanoscale optical cavity},}\ }\href@noop
  {} {\bibfield  {journal} {\bibinfo  {journal} {Science}\ }\textbf {\bibinfo
  {volume} {340}},\ \bibinfo {pages} {1202--1205} (\bibinfo {year}
  {2013})}\BibitemShut {NoStop}%
\bibitem [{\citenamefont {Goban}\ \emph {et~al.}(2015)\citenamefont {Goban},
  \citenamefont {Hung}, \citenamefont {Hood}, \citenamefont {Yu}, \citenamefont
  {Muniz}, \citenamefont {Painter},\ and\ \citenamefont
  {Kimble}}]{goban2015superradiance}%
  \BibitemOpen
  \bibfield  {author} {\bibinfo {author} {\bibfnamefont {A}~\bibnamefont
  {Goban}}, \bibinfo {author} {\bibfnamefont {C-L}\ \bibnamefont {Hung}},
  \bibinfo {author} {\bibfnamefont {JD}~\bibnamefont {Hood}}, \bibinfo {author}
  {\bibfnamefont {S-P}\ \bibnamefont {Yu}}, \bibinfo {author} {\bibfnamefont
  {JA}~\bibnamefont {Muniz}}, \bibinfo {author} {\bibfnamefont {O}~\bibnamefont
  {Painter}}, \ and\ \bibinfo {author} {\bibfnamefont {HJ}~\bibnamefont
  {Kimble}},\ }\bibfield  {title} {\enquote {\bibinfo {title} {Superradiance
  for atoms trapped along a photonic crystal waveguide},}\ }\href@noop {}
  {\bibfield  {journal} {\bibinfo  {journal} {Physical review letters}\
  }\textbf {\bibinfo {volume} {115}},\ \bibinfo {pages} {063601} (\bibinfo
  {year} {2015})}\BibitemShut {NoStop}%
\bibitem [{\citenamefont {Urban}\ \emph {et~al.}(2009)\citenamefont {Urban},
  \citenamefont {Johnson}, \citenamefont {Henage}, \citenamefont {Isenhower},
  \citenamefont {Yavuz}, \citenamefont {Walker},\ and\ \citenamefont
  {Saffman}}]{Urban2009}%
  \BibitemOpen
  \bibfield  {author} {\bibinfo {author} {\bibfnamefont {E.}~\bibnamefont
  {Urban}}, \bibinfo {author} {\bibfnamefont {T.~A.}\ \bibnamefont {Johnson}},
  \bibinfo {author} {\bibfnamefont {T.}~\bibnamefont {Henage}}, \bibinfo
  {author} {\bibfnamefont {L.}~\bibnamefont {Isenhower}}, \bibinfo {author}
  {\bibfnamefont {D.~D.}\ \bibnamefont {Yavuz}}, \bibinfo {author}
  {\bibfnamefont {T.~G.}\ \bibnamefont {Walker}}, \ and\ \bibinfo {author}
  {\bibfnamefont {M.}~\bibnamefont {Saffman}},\ }\bibfield  {title} {\enquote
  {\bibinfo {title} {Observation of rydberg blockade between two atoms},}\
  }\href {\doibase 10.1038/nphys1178} {\bibfield  {journal} {\bibinfo
  {journal} {Nat. Phys.}\ }\textbf {\bibinfo {volume} {5}},\ \bibinfo {pages}
  {110--114} (\bibinfo {year} {2009})}\BibitemShut {NoStop}%
\bibitem [{\citenamefont {Ga\"etan}\ \emph {et~al.}(2009)\citenamefont
  {Ga\"etan}, \citenamefont {Miroshnychenko}, \citenamefont {Wilk},
  \citenamefont {Chotia}, \citenamefont {Viteau}, \citenamefont {Comparat},
  \citenamefont {Pillet}, \citenamefont {Browaeys},\ and\ \citenamefont
  {Grangier}}]{Gaetan2009}%
  \BibitemOpen
  \bibfield  {author} {\bibinfo {author} {\bibfnamefont {A.}~\bibnamefont
  {Ga\"etan}}, \bibinfo {author} {\bibfnamefont {Y.}~\bibnamefont
  {Miroshnychenko}}, \bibinfo {author} {\bibfnamefont {T.}~\bibnamefont
  {Wilk}}, \bibinfo {author} {\bibfnamefont {A.}~\bibnamefont {Chotia}},
  \bibinfo {author} {\bibfnamefont {M.}~\bibnamefont {Viteau}}, \bibinfo
  {author} {\bibfnamefont {D.}~\bibnamefont {Comparat}}, \bibinfo {author}
  {\bibfnamefont {P.}~\bibnamefont {Pillet}}, \bibinfo {author} {\bibfnamefont
  {A.}~\bibnamefont {Browaeys}}, \ and\ \bibinfo {author} {\bibfnamefont
  {P.}~\bibnamefont {Grangier}},\ }\bibfield  {title} {\enquote {\bibinfo
  {title} {Observation of collective excitation of two individual atoms in the
  rydberg blockade regime},}\ }\href {\doibase 10.1038/nphys1183} {\bibfield
  {journal} {\bibinfo  {journal} {Nat. Phys.}\ }\textbf {\bibinfo {volume}
  {5}},\ \bibinfo {pages} {115--118} (\bibinfo {year} {2009})}\BibitemShut
  {NoStop}%
\bibitem [{\citenamefont {Dudin}\ \emph {et~al.}(2012)\citenamefont {Dudin},
  \citenamefont {Li}, \citenamefont {Bariani},\ and\ \citenamefont
  {Kuzmich}}]{dudin2012observation}%
  \BibitemOpen
  \bibfield  {author} {\bibinfo {author} {\bibfnamefont {YO}~\bibnamefont
  {Dudin}}, \bibinfo {author} {\bibfnamefont {L}~\bibnamefont {Li}}, \bibinfo
  {author} {\bibfnamefont {F}~\bibnamefont {Bariani}}, \ and\ \bibinfo {author}
  {\bibfnamefont {A}~\bibnamefont {Kuzmich}},\ }\bibfield  {title} {\enquote
  {\bibinfo {title} {Observation of coherent many-body rabi oscillations},}\
  }\href@noop {} {\bibfield  {journal} {\bibinfo  {journal} {Nature Physics}\
  }\textbf {\bibinfo {volume} {8}},\ \bibinfo {pages} {790--794} (\bibinfo
  {year} {2012})}\BibitemShut {NoStop}%
\bibitem [{\citenamefont {Dudin}\ and\ \citenamefont
  {Kuzmich}(2012)}]{dudi2012stro}%
  \BibitemOpen
  \bibfield  {author} {\bibinfo {author} {\bibfnamefont {Y.~O.}\ \bibnamefont
  {Dudin}}\ and\ \bibinfo {author} {\bibfnamefont {A.}~\bibnamefont
  {Kuzmich}},\ }\bibfield  {title} {\enquote {\bibinfo {title} {Strongly
  interacting {Rydberg} excitations of a cold atomic gas},}\ }\href@noop {}
  {\bibfield  {journal} {\bibinfo  {journal} {Science}\ }\textbf {\bibinfo
  {volume} {336}},\ \bibinfo {pages} {887--889} (\bibinfo {year}
  {2012})}\BibitemShut {NoStop}%
\bibitem [{\citenamefont {Tiarks}\ \emph {et~al.}(2014)\citenamefont {Tiarks},
  \citenamefont {Baur}, \citenamefont {Schneider}, \citenamefont {D\"urr},\
  and\ \citenamefont {Rempe}}]{tiar2014sing}%
  \BibitemOpen
  \bibfield  {author} {\bibinfo {author} {\bibfnamefont {D.}~\bibnamefont
  {Tiarks}}, \bibinfo {author} {\bibfnamefont {S.}~\bibnamefont {Baur}},
  \bibinfo {author} {\bibfnamefont {K.}~\bibnamefont {Schneider}}, \bibinfo
  {author} {\bibfnamefont {S.}~\bibnamefont {D\"urr}}, \ and\ \bibinfo {author}
  {\bibfnamefont {G.}~\bibnamefont {Rempe}},\ }\bibfield  {title} {\enquote
  {\bibinfo {title} {Single-{Photon} {Transistor} {Using} a {F\"orster}
  {Resonance}},}\ }\href {\doibase 10.1103/PhysRevLett.113.053602} {\bibfield
  {journal} {\bibinfo  {journal} {Phys. Rev. Lett.}\ }\textbf {\bibinfo
  {volume} {113}},\ \bibinfo {pages} {053602} (\bibinfo {year}
  {2014})}\BibitemShut {NoStop}%
\bibitem [{\citenamefont {Gorniaczyk}\ \emph {et~al.}(2014)\citenamefont
  {Gorniaczyk}, \citenamefont {Tresp}, \citenamefont {Schmidt}, \citenamefont
  {Fedder},\ and\ \citenamefont {Hofferberth}}]{gorn2014sing}%
  \BibitemOpen
  \bibfield  {author} {\bibinfo {author} {\bibfnamefont {H.}~\bibnamefont
  {Gorniaczyk}}, \bibinfo {author} {\bibfnamefont {C.}~\bibnamefont {Tresp}},
  \bibinfo {author} {\bibfnamefont {J.}~\bibnamefont {Schmidt}}, \bibinfo
  {author} {\bibfnamefont {H.}~\bibnamefont {Fedder}}, \ and\ \bibinfo {author}
  {\bibfnamefont {S.}~\bibnamefont {Hofferberth}},\ }\bibfield  {title}
  {\enquote {\bibinfo {title} {Single-{Photon} {Transistor} {Mediated} by
  {Interstate} {Rydberg} {Interactions}},}\ }\href {\doibase
  10.1103/PhysRevLett.113.053601} {\bibfield  {journal} {\bibinfo  {journal}
  {Phys. Rev. Lett.}\ }\textbf {\bibinfo {volume} {113}},\ \bibinfo {pages}
  {053601} (\bibinfo {year} {2014})}\BibitemShut {NoStop}%
\bibitem [{\citenamefont {Guerlin}\ \emph {et~al.}(2010)\citenamefont
  {Guerlin}, \citenamefont {Brion}, \citenamefont {Esslinger},\ and\
  \citenamefont {M\o{}lmer}}]{Guerlin2010}%
  \BibitemOpen
  \bibfield  {author} {\bibinfo {author} {\bibfnamefont {Christine}\
  \bibnamefont {Guerlin}}, \bibinfo {author} {\bibfnamefont {Etienne}\
  \bibnamefont {Brion}}, \bibinfo {author} {\bibfnamefont {Tilman}\
  \bibnamefont {Esslinger}}, \ and\ \bibinfo {author} {\bibfnamefont {Klaus}\
  \bibnamefont {M\o{}lmer}},\ }\bibfield  {title} {\enquote {\bibinfo {title}
  {Cavity quantum electrodynamics with a rydberg-blocked atomic ensemble},}\
  }\href {\doibase 10.1103/PhysRevA.82.053832} {\bibfield  {journal} {\bibinfo
  {journal} {Phys. Rev. A}\ }\textbf {\bibinfo {volume} {82}},\ \bibinfo
  {pages} {053832} (\bibinfo {year} {2010})}\BibitemShut {NoStop}%
\bibitem [{\citenamefont {Parigi}\ \emph {et~al.}(2012)\citenamefont {Parigi},
  \citenamefont {Bimbard}, \citenamefont {Stanojevic}, \citenamefont
  {Hilliard}, \citenamefont {Nogrette}, \citenamefont {Tualle-Brouri},
  \citenamefont {Ourjoumtsev},\ and\ \citenamefont {Grangier}}]{pari2012obse}%
  \BibitemOpen
  \bibfield  {author} {\bibinfo {author} {\bibfnamefont {Valentina}\
  \bibnamefont {Parigi}}, \bibinfo {author} {\bibfnamefont {Erwan}\
  \bibnamefont {Bimbard}}, \bibinfo {author} {\bibfnamefont {Jovica}\
  \bibnamefont {Stanojevic}}, \bibinfo {author} {\bibfnamefont {Andrew~J.}\
  \bibnamefont {Hilliard}}, \bibinfo {author} {\bibfnamefont {Florence}\
  \bibnamefont {Nogrette}}, \bibinfo {author} {\bibfnamefont {Rosa}\
  \bibnamefont {Tualle-Brouri}}, \bibinfo {author} {\bibfnamefont {Alexei}\
  \bibnamefont {Ourjoumtsev}}, \ and\ \bibinfo {author} {\bibfnamefont
  {Philippe}\ \bibnamefont {Grangier}},\ }\bibfield  {title} {\enquote
  {\bibinfo {title} {Observation and {Measurement} of {Interaction}-{Induced}
  {Dispersive} {Optical} {Nonlinearities} in an {Ensemble} of {Cold} {Rydberg}
  {Atoms}},}\ }\href {\doibase 10.1103/PhysRevLett.109.233602} {\bibfield
  {journal} {\bibinfo  {journal} {Phys. Rev. Lett.}\ }\textbf {\bibinfo
  {volume} {109}},\ \bibinfo {pages} {233602} (\bibinfo {year}
  {2012})}\BibitemShut {NoStop}%
\bibitem [{\citenamefont {Ningyuan}\ \emph {et~al.}(2016)\citenamefont
  {Ningyuan}, \citenamefont {Georgakopoulos}, \citenamefont {Ryou},
  \citenamefont {Schine}, \citenamefont {Sommer},\ and\ \citenamefont
  {Simon}}]{ningyuan2016observation}%
  \BibitemOpen
  \bibfield  {author} {\bibinfo {author} {\bibfnamefont {Jia}\ \bibnamefont
  {Ningyuan}}, \bibinfo {author} {\bibfnamefont {Alexandros}\ \bibnamefont
  {Georgakopoulos}}, \bibinfo {author} {\bibfnamefont {Albert}\ \bibnamefont
  {Ryou}}, \bibinfo {author} {\bibfnamefont {Nathan}\ \bibnamefont {Schine}},
  \bibinfo {author} {\bibfnamefont {Ariel}\ \bibnamefont {Sommer}}, \ and\
  \bibinfo {author} {\bibfnamefont {Jonathan}\ \bibnamefont {Simon}},\
  }\bibfield  {title} {\enquote {\bibinfo {title} {Observation and
  characterization of cavity rydberg polaritons},}\ }\href@noop {} {\bibfield
  {journal} {\bibinfo  {journal} {Physical Review A}\ }\textbf {\bibinfo
  {volume} {93}},\ \bibinfo {pages} {041802} (\bibinfo {year}
  {2016})}\BibitemShut {NoStop}%
\bibitem [{\citenamefont {Anderson}\ \emph {et~al.}(2016)\citenamefont
  {Anderson}, \citenamefont {Ma}, \citenamefont {Owens}, \citenamefont
  {Schuster},\ and\ \citenamefont {Simon}}]{anderson2016engineering}%
  \BibitemOpen
  \bibfield  {author} {\bibinfo {author} {\bibfnamefont {Brandon~M}\
  \bibnamefont {Anderson}}, \bibinfo {author} {\bibfnamefont {Ruichao}\
  \bibnamefont {Ma}}, \bibinfo {author} {\bibfnamefont {Clai}\ \bibnamefont
  {Owens}}, \bibinfo {author} {\bibfnamefont {David~I}\ \bibnamefont
  {Schuster}}, \ and\ \bibinfo {author} {\bibfnamefont {Jonathan}\ \bibnamefont
  {Simon}},\ }\bibfield  {title} {\enquote {\bibinfo {title} {Engineering
  topological many-body materials in microwave cavity arrays},}\ }\href@noop {}
  {\bibfield  {journal} {\bibinfo  {journal} {Physical Review X}\ }\textbf
  {\bibinfo {volume} {6}},\ \bibinfo {pages} {041043} (\bibinfo {year}
  {2016})}\BibitemShut {NoStop}%
\bibitem [{\citenamefont {Gorshkov}\ \emph {et~al.}(2011)\citenamefont
  {Gorshkov}, \citenamefont {Otterbach}, \citenamefont {Fleischhauer},
  \citenamefont {Pohl},\ and\ \citenamefont {Lukin}}]{gorshkov2011photon}%
  \BibitemOpen
  \bibfield  {author} {\bibinfo {author} {\bibfnamefont {Alexey~V}\
  \bibnamefont {Gorshkov}}, \bibinfo {author} {\bibfnamefont {Johannes}\
  \bibnamefont {Otterbach}}, \bibinfo {author} {\bibfnamefont {Michael}\
  \bibnamefont {Fleischhauer}}, \bibinfo {author} {\bibfnamefont {Thomas}\
  \bibnamefont {Pohl}}, \ and\ \bibinfo {author} {\bibfnamefont {Mikhail~D}\
  \bibnamefont {Lukin}},\ }\bibfield  {title} {\enquote {\bibinfo {title}
  {Photon-photon interactions via rydberg blockade},}\ }\href@noop {}
  {\bibfield  {journal} {\bibinfo  {journal} {Physical review letters}\
  }\textbf {\bibinfo {volume} {107}},\ \bibinfo {pages} {133602} (\bibinfo
  {year} {2011})}\BibitemShut {NoStop}%
\bibitem [{\citenamefont {Baumann}\ \emph {et~al.}(2010)\citenamefont
  {Baumann}, \citenamefont {Guerlin}, \citenamefont {Brennecke},\ and\
  \citenamefont {Esslinger}}]{baumann2010dicke}%
  \BibitemOpen
  \bibfield  {author} {\bibinfo {author} {\bibfnamefont {Kristian}\
  \bibnamefont {Baumann}}, \bibinfo {author} {\bibfnamefont {Christine}\
  \bibnamefont {Guerlin}}, \bibinfo {author} {\bibfnamefont {Ferdinand}\
  \bibnamefont {Brennecke}}, \ and\ \bibinfo {author} {\bibfnamefont {Tilman}\
  \bibnamefont {Esslinger}},\ }\bibfield  {title} {\enquote {\bibinfo {title}
  {Dicke quantum phase transition with a superfluid gas in an optical
  cavity},}\ }\href@noop {} {\bibfield  {journal} {\bibinfo  {journal}
  {Nature}\ }\textbf {\bibinfo {volume} {464}},\ \bibinfo {pages} {1301--1306}
  (\bibinfo {year} {2010})}\BibitemShut {NoStop}%
\bibitem [{\citenamefont {L{\'e}onard}\ \emph {et~al.}(2016)\citenamefont
  {L{\'e}onard}, \citenamefont {Morales}, \citenamefont {Zupancic},
  \citenamefont {Esslinger},\ and\ \citenamefont
  {Donner}}]{leonard2016supersolid}%
  \BibitemOpen
  \bibfield  {author} {\bibinfo {author} {\bibfnamefont {Julian}\ \bibnamefont
  {L{\'e}onard}}, \bibinfo {author} {\bibfnamefont {Andrea}\ \bibnamefont
  {Morales}}, \bibinfo {author} {\bibfnamefont {Philip}\ \bibnamefont
  {Zupancic}}, \bibinfo {author} {\bibfnamefont {Tilman}\ \bibnamefont
  {Esslinger}}, \ and\ \bibinfo {author} {\bibfnamefont {Tobias}\ \bibnamefont
  {Donner}},\ }\bibfield  {title} {\enquote {\bibinfo {title} {Supersolid
  formation in a quantum gas breaking continuous translational symmetry},}\
  }\href@noop {} {\bibfield  {journal} {\bibinfo  {journal} {Nature}\ }\textbf
  {\bibinfo {volume} {543}},\ \bibinfo {pages} {87} (\bibinfo {year}
  {2016})}\BibitemShut {NoStop}%
\bibitem [{\citenamefont {Klaers}\ \emph {et~al.}(2010)\citenamefont {Klaers},
  \citenamefont {Schmitt}, \citenamefont {Vewinger},\ and\ \citenamefont
  {Weitz}}]{klaers2010bose}%
  \BibitemOpen
  \bibfield  {author} {\bibinfo {author} {\bibfnamefont {Jan}\ \bibnamefont
  {Klaers}}, \bibinfo {author} {\bibfnamefont {Julian}\ \bibnamefont
  {Schmitt}}, \bibinfo {author} {\bibfnamefont {Frank}\ \bibnamefont
  {Vewinger}}, \ and\ \bibinfo {author} {\bibfnamefont {Martin}\ \bibnamefont
  {Weitz}},\ }\bibfield  {title} {\enquote {\bibinfo {title} {Bose-einstein
  condensation of photons in an optical microcavity},}\ }\href@noop {}
  {\bibfield  {journal} {\bibinfo  {journal} {Nature}\ }\textbf {\bibinfo
  {volume} {468}},\ \bibinfo {pages} {545--548} (\bibinfo {year}
  {2010})}\BibitemShut {NoStop}%
\bibitem [{\citenamefont {Raftery}\ \emph {et~al.}(2014)\citenamefont
  {Raftery}, \citenamefont {Sadri}, \citenamefont {Schmidt}, \citenamefont
  {T{\"u}reci},\ and\ \citenamefont {Houck}}]{raftery2014observation}%
  \BibitemOpen
  \bibfield  {author} {\bibinfo {author} {\bibfnamefont {James}\ \bibnamefont
  {Raftery}}, \bibinfo {author} {\bibfnamefont {Darius}\ \bibnamefont {Sadri}},
  \bibinfo {author} {\bibfnamefont {Sebastian}\ \bibnamefont {Schmidt}},
  \bibinfo {author} {\bibfnamefont {Hakan~E}\ \bibnamefont {T{\"u}reci}}, \
  and\ \bibinfo {author} {\bibfnamefont {Andrew~A}\ \bibnamefont {Houck}},\
  }\bibfield  {title} {\enquote {\bibinfo {title} {Observation of a
  dissipation-induced classical to quantum transition},}\ }\href@noop {}
  {\bibfield  {journal} {\bibinfo  {journal} {Physical Review X}\ }\textbf
  {\bibinfo {volume} {4}},\ \bibinfo {pages} {031043} (\bibinfo {year}
  {2014})}\BibitemShut {NoStop}%
\bibitem [{\citenamefont {Hafezi}\ \emph {et~al.}(2015)\citenamefont {Hafezi},
  \citenamefont {Adhikari},\ and\ \citenamefont {Taylor}}]{hafezi2015}%
  \BibitemOpen
  \bibfield  {author} {\bibinfo {author} {\bibfnamefont {M}~\bibnamefont
  {Hafezi}}, \bibinfo {author} {\bibfnamefont {P}~\bibnamefont {Adhikari}}, \
  and\ \bibinfo {author} {\bibfnamefont {JM}~\bibnamefont {Taylor}},\
  }\bibfield  {title} {\enquote {\bibinfo {title} {Chemical potential for light
  by parametric coupling},}\ }\href@noop {} {\bibfield  {journal} {\bibinfo
  {journal} {Physical Review B}\ }\textbf {\bibinfo {volume} {92}},\ \bibinfo
  {pages} {174305} (\bibinfo {year} {2015})}\BibitemShut {NoStop}%
\bibitem [{\citenamefont {Lebreuilly}\ \emph {et~al.}(2016)\citenamefont
  {Lebreuilly}, \citenamefont {Wouters},\ and\ \citenamefont
  {Carusotto}}]{lebreuilly2016towards}%
  \BibitemOpen
  \bibfield  {author} {\bibinfo {author} {\bibfnamefont {Jos{\'e}}\
  \bibnamefont {Lebreuilly}}, \bibinfo {author} {\bibfnamefont {Michiel}\
  \bibnamefont {Wouters}}, \ and\ \bibinfo {author} {\bibfnamefont {Iacopo}\
  \bibnamefont {Carusotto}},\ }\bibfield  {title} {\enquote {\bibinfo {title}
  {Towards strongly correlated photons in arrays of dissipative nonlinear
  cavities under a frequency-dependent incoherent pumping},}\ }\href@noop {}
  {\bibfield  {journal} {\bibinfo  {journal} {Comptes Rendus Physique}\
  }\textbf {\bibinfo {volume} {17}},\ \bibinfo {pages} {836--860} (\bibinfo
  {year} {2016})}\BibitemShut {NoStop}%
\bibitem [{\citenamefont {Ma}\ \emph {et~al.}(2017)\citenamefont {Ma},
  \citenamefont {Owens}, \citenamefont {Houck}, \citenamefont {Schuster},\ and\
  \citenamefont {Simon}}]{ma2017autonomous}%
  \BibitemOpen
  \bibfield  {author} {\bibinfo {author} {\bibfnamefont {Ruichao}\ \bibnamefont
  {Ma}}, \bibinfo {author} {\bibfnamefont {Clai}\ \bibnamefont {Owens}},
  \bibinfo {author} {\bibfnamefont {Andrew}\ \bibnamefont {Houck}}, \bibinfo
  {author} {\bibfnamefont {David~I.}\ \bibnamefont {Schuster}}, \ and\ \bibinfo
  {author} {\bibfnamefont {Jonathan}\ \bibnamefont {Simon}},\ }\bibfield
  {title} {\enquote {\bibinfo {title} {Autonomous stabilizer for incompressible
  photon fluids and solids},}\ }\href {\doibase 10.1103/PhysRevA.95.043811}
  {\bibfield  {journal} {\bibinfo  {journal} {Phys. Rev. A}\ }\textbf {\bibinfo
  {volume} {95}},\ \bibinfo {pages} {043811} (\bibinfo {year}
  {2017})}\BibitemShut {NoStop}%
\bibitem [{\citenamefont {Kouwenhoven}\ \emph {et~al.}(1997)\citenamefont
  {Kouwenhoven}, \citenamefont {Marcus}, \citenamefont {McEuen}, \citenamefont
  {Tarucha}, \citenamefont {Westervelt},\ and\ \citenamefont
  {Wingreen}}]{kouwenhoven1997electron}%
  \BibitemOpen
  \bibfield  {author} {\bibinfo {author} {\bibfnamefont {Leo~P}\ \bibnamefont
  {Kouwenhoven}}, \bibinfo {author} {\bibfnamefont {Charles~M}\ \bibnamefont
  {Marcus}}, \bibinfo {author} {\bibfnamefont {Paul~L}\ \bibnamefont {McEuen}},
  \bibinfo {author} {\bibfnamefont {Seigo}\ \bibnamefont {Tarucha}}, \bibinfo
  {author} {\bibfnamefont {Robert~M}\ \bibnamefont {Westervelt}}, \ and\
  \bibinfo {author} {\bibfnamefont {Ned~S}\ \bibnamefont {Wingreen}},\
  }\bibfield  {title} {\enquote {\bibinfo {title} {Electron transport in
  quantum dots},}\ }in\ \href@noop {} {\emph {\bibinfo {booktitle} {Mesoscopic
  electron transport}}}\ (\bibinfo  {publisher} {Springer},\ \bibinfo {year}
  {1997})\ pp.\ \bibinfo {pages} {105--214}\BibitemShut {NoStop}%
\bibitem [{\citenamefont {Fleischhauer}\ \emph {et~al.}(2005)\citenamefont
  {Fleischhauer}, \citenamefont {Imamoglu},\ and\ \citenamefont
  {Marangos}}]{fleischhauer2005electromagnetically}%
  \BibitemOpen
  \bibfield  {author} {\bibinfo {author} {\bibfnamefont {Michael}\ \bibnamefont
  {Fleischhauer}}, \bibinfo {author} {\bibfnamefont {Atac}\ \bibnamefont
  {Imamoglu}}, \ and\ \bibinfo {author} {\bibfnamefont {Jonathan~P}\
  \bibnamefont {Marangos}},\ }\bibfield  {title} {\enquote {\bibinfo {title}
  {Electromagnetically induced transparency: Optics in coherent media},}\
  }\href@noop {} {\bibfield  {journal} {\bibinfo  {journal} {Reviews of modern
  physics}\ }\textbf {\bibinfo {volume} {77}},\ \bibinfo {pages} {633}
  (\bibinfo {year} {2005})}\BibitemShut {NoStop}%
\bibitem [{\citenamefont {Zhao}\ \emph {et~al.}(2010)\citenamefont {Zhao},
  \citenamefont {M{\"u}ller}, \citenamefont {Hammerer},\ and\ \citenamefont
  {Zoller}}]{zhao2010efficient}%
  \BibitemOpen
  \bibfield  {author} {\bibinfo {author} {\bibfnamefont {Bo}~\bibnamefont
  {Zhao}}, \bibinfo {author} {\bibfnamefont {Markus}\ \bibnamefont
  {M{\"u}ller}}, \bibinfo {author} {\bibfnamefont {Klemens}\ \bibnamefont
  {Hammerer}}, \ and\ \bibinfo {author} {\bibfnamefont {Peter}\ \bibnamefont
  {Zoller}},\ }\bibfield  {title} {\enquote {\bibinfo {title} {Efficient
  quantum repeater based on deterministic rydberg gates},}\ }\href@noop {}
  {\bibfield  {journal} {\bibinfo  {journal} {Physical Review A}\ }\textbf
  {\bibinfo {volume} {81}},\ \bibinfo {pages} {052329} (\bibinfo {year}
  {2010})}\BibitemShut {NoStop}%
\bibitem [{\citenamefont {Brion}\ \emph {et~al.}(2012)\citenamefont {Brion},
  \citenamefont {Carlier}, \citenamefont {Akulin},\ and\ \citenamefont
  {M{\o}lmer}}]{brion2012quantum}%
  \BibitemOpen
  \bibfield  {author} {\bibinfo {author} {\bibfnamefont {Etienne}\ \bibnamefont
  {Brion}}, \bibinfo {author} {\bibfnamefont {F}~\bibnamefont {Carlier}},
  \bibinfo {author} {\bibfnamefont {VM}~\bibnamefont {Akulin}}, \ and\ \bibinfo
  {author} {\bibfnamefont {Klaus}\ \bibnamefont {M{\o}lmer}},\ }\bibfield
  {title} {\enquote {\bibinfo {title} {Quantum repeater with rydberg-blocked
  atomic ensembles in fiber-coupled cavities},}\ }\href@noop {} {\bibfield
  {journal} {\bibinfo  {journal} {Physical Review A}\ }\textbf {\bibinfo
  {volume} {85}},\ \bibinfo {pages} {042324} (\bibinfo {year}
  {2012})}\BibitemShut {NoStop}%
\bibitem [{\citenamefont {Bienias}(2016)}]{bienias2016few}%
  \BibitemOpen
  \bibfield  {author} {\bibinfo {author} {\bibfnamefont {Przemyslaw}\
  \bibnamefont {Bienias}},\ }\bibfield  {title} {\enquote {\bibinfo {title}
  {Few-body quantum physics with strongly interacting rydberg polaritons},}\
  }\href@noop {} {\bibfield  {journal} {\bibinfo  {journal} {The European
  Physical Journal Special Topics}\ }\textbf {\bibinfo {volume} {225}},\
  \bibinfo {pages} {2957--2976} (\bibinfo {year} {2016})}\BibitemShut {NoStop}%
\bibitem [{\citenamefont {Grusdt}\ and\ \citenamefont
  {Fleischhauer}(2013)}]{grusdt2013fractional}%
  \BibitemOpen
  \bibfield  {author} {\bibinfo {author} {\bibfnamefont {Fabian}\ \bibnamefont
  {Grusdt}}\ and\ \bibinfo {author} {\bibfnamefont {Michael}\ \bibnamefont
  {Fleischhauer}},\ }\bibfield  {title} {\enquote {\bibinfo {title} {Fractional
  quantum hall physics with ultracold rydberg gases in artificial gauge
  fields},}\ }\href@noop {} {\bibfield  {journal} {\bibinfo  {journal}
  {Physical Review A}\ }\textbf {\bibinfo {volume} {87}},\ \bibinfo {pages}
  {043628} (\bibinfo {year} {2013})}\BibitemShut {NoStop}%
\bibitem [{\citenamefont {Grusdt}\ \emph {et~al.}(2016)\citenamefont {Grusdt},
  \citenamefont {Yao}, \citenamefont {Abanin}, \citenamefont {Fleischhauer},\
  and\ \citenamefont {Demler}}]{grusdt2016interferometric}%
  \BibitemOpen
  \bibfield  {author} {\bibinfo {author} {\bibfnamefont {Fabian}\ \bibnamefont
  {Grusdt}}, \bibinfo {author} {\bibfnamefont {Norman~Y}\ \bibnamefont {Yao}},
  \bibinfo {author} {\bibfnamefont {D}~\bibnamefont {Abanin}}, \bibinfo
  {author} {\bibfnamefont {Michael}\ \bibnamefont {Fleischhauer}}, \ and\
  \bibinfo {author} {\bibfnamefont {E}~\bibnamefont {Demler}},\ }\bibfield
  {title} {\enquote {\bibinfo {title} {Interferometric measurements of
  many-body topological invariants using mobile impurities},}\ }\href@noop {}
  {\bibfield  {journal} {\bibinfo  {journal} {Nature Communications}\ }\textbf
  {\bibinfo {volume} {7}} (\bibinfo {year} {2016})}\BibitemShut {NoStop}%
\bibitem [{\citenamefont {Umucal{\i}lar}\ and\ \citenamefont
  {Carusotto}(2013)}]{umucalilar2013many}%
  \BibitemOpen
  \bibfield  {author} {\bibinfo {author} {\bibfnamefont {RO}~\bibnamefont
  {Umucal{\i}lar}}\ and\ \bibinfo {author} {\bibfnamefont {I}~\bibnamefont
  {Carusotto}},\ }\bibfield  {title} {\enquote {\bibinfo {title} {Many-body
  braiding phases in a rotating strongly correlated photon gas},}\ }\href@noop
  {} {\bibfield  {journal} {\bibinfo  {journal} {Physics Letters A}\ }\textbf
  {\bibinfo {volume} {377}},\ \bibinfo {pages} {2074--2078} (\bibinfo {year}
  {2013})}\BibitemShut {NoStop}%
\bibitem [{\citenamefont {{Tanji-Suzuki}}\ \emph {et~al.}(2011)\citenamefont
  {{Tanji-Suzuki}}, \citenamefont {{Leroux}}, \citenamefont {{Schleier-Smith}},
  \citenamefont {{Cetina}}, \citenamefont {{Grier}}, \citenamefont {{Simon}},\
  and\ \citenamefont {{Vuleti{\'c}}}}]{Tanji2011b}%
  \BibitemOpen
  \bibfield  {author} {\bibinfo {author} {\bibfnamefont {H.}~\bibnamefont
  {{Tanji-Suzuki}}}, \bibinfo {author} {\bibfnamefont {I.~D.}\ \bibnamefont
  {{Leroux}}}, \bibinfo {author} {\bibfnamefont {M.~H.}\ \bibnamefont
  {{Schleier-Smith}}}, \bibinfo {author} {\bibfnamefont {M.}~\bibnamefont
  {{Cetina}}}, \bibinfo {author} {\bibfnamefont {A.~T.}\ \bibnamefont
  {{Grier}}}, \bibinfo {author} {\bibfnamefont {J.}~\bibnamefont {{Simon}}}, \
  and\ \bibinfo {author} {\bibfnamefont {V.}~\bibnamefont {{Vuleti{\'c}}}},\
  }\bibfield  {title} {\enquote {\bibinfo {title} {{Interaction between Atomic
  Ensembles and Optical Resonators}},}\ }\href {\doibase
  10.1016/B978-0-12-385508-4.00004-8} {\bibfield  {journal} {\bibinfo
  {journal} {Advances in Atomic Molecular and Optical Physics}\ }\textbf
  {\bibinfo {volume} {60}},\ \bibinfo {pages} {201--237} (\bibinfo {year}
  {2011})}\BibitemShut {NoStop}%
\bibitem [{\citenamefont {Mirgorodskiy}\ \emph {et~al.}(2017)\citenamefont
  {Mirgorodskiy}, \citenamefont {Christaller}, \citenamefont {Braun},
  \citenamefont {Paris-Mandoki}, \citenamefont {Tresp},\ and\ \citenamefont
  {Hofferberth}}]{mirgorodskiy2017electromagnetically}%
  \BibitemOpen
  \bibfield  {author} {\bibinfo {author} {\bibfnamefont {Ivan}\ \bibnamefont
  {Mirgorodskiy}}, \bibinfo {author} {\bibfnamefont {Florian}\ \bibnamefont
  {Christaller}}, \bibinfo {author} {\bibfnamefont {Christoph}\ \bibnamefont
  {Braun}}, \bibinfo {author} {\bibfnamefont {Asaf}\ \bibnamefont
  {Paris-Mandoki}}, \bibinfo {author} {\bibfnamefont {Christoph}\ \bibnamefont
  {Tresp}}, \ and\ \bibinfo {author} {\bibfnamefont {Sebastian}\ \bibnamefont
  {Hofferberth}},\ }\bibfield  {title} {\enquote {\bibinfo {title}
  {Electromagnetically induced transparency of ultralong-range rydberg
  molecules},}\ }\href@noop {} {\bibfield  {journal} {\bibinfo  {journal}
  {arXiv preprint arXiv:1705.03700}\ } (\bibinfo {year} {2017})}\BibitemShut
  {NoStop}%
\end{thebibliography}%
\incltocpage

\newpage

\tableofcontents

\appendix
\setcounter{secnumdepth}{1}
\section{Methods}
\label{sec:Methods}
Our experiments begin with a magneto-optical trap (MOT) of $10^{7}$ $^{87}$Rb atoms which is polarization-gradient-cooled to a temperature of 15 $\mu$K, and loaded into a 1D (vertical) optical conveyor belt (with waist 85 $\mu$m). It is then transported $32.1$ mm in 13 ms into a four-mirror optical resonator, by detuning one of the lattice beams by up to 5 MHz, with a maximum acceleration of $\sim 1000$m/s$^2$. The resonator has a waist of $12\mu$m$\times14\mu$m, located at the position of the atomic sample, and a finesse of $\mathcal{F}=1480(50)$ at $780$nm, on the Rb D2 line (see SI \ref{SI:CavityDetails}). This results in a maximal single-atom cooperativity on the cavity axis of $\eta=0.26$ \cite{Tanji2011b}, corresponding resonator linewidth $\kappa=2\pi\times 1.6$ MHz and single-atom single-quantum Rabi frequency $g=0.78$ MHz on the $|F=2,m_F=2\rangle\rightarrow|3',3\rangle$ transition of the $^{87}$Rb D2 line, on the resonator axis, for a circularly-polarized TEM$_{00}$ running-wave mode. Although the cavity mode is linearly polarized, using the maximal Clebsch-Gordan coefficient provides a well-defined notion of effective atom number.

The atomic cloud at the location of the resonator has dimensions $35\mu$m$\times41\mu$m$\times37\mu$m, at a density of $1.1\times 10^{11} $cm$^{-3}$; to realize a strongly interacting 0-D quantum dot, the sample must be smaller than the blockade radius of $15\mu$m in the 100S Rydberg state~\cite{Saffman_2010}. The optical resonator waist defines a $\frac{1}{e^2}$ sample of radius $\sim 13\mu$m along two axes; unlike free-space experiments~\cite{peyronel2012quantum}, our 0-D resonator-defined dot must be smaller than a blockade radius even along the resonator axis in order to enter the strongly interacting regime. To achieve this longitudinal confinement, we employ super-resolution slicing: the atoms are locally re-pumped into the $|F=2\rangle$ hyperfine state (which couples strongly to the resonator mode) with a beam whose waist is 81 $\mu$m, and then depumped with a $\sim$ TEM$_{10}$ beam whose line-node is located within the atomic cloud, leaving a sample with RMS length $\approx 10\mu$m (see SI ~\ref{SI:AtomSlicing}).

To capture photon arrival times for correlation experiments, we employ a home-built 8-channel photon time-tagger with 8 ns resolution on each channel, based upon an Opal Kelly XEM6001 field-programmable gate array. A similar device is employed to store histogram data for ring-down and spectroscopy experiments.

\section{Atom Slicing\label{SI:AtomSlicing}}
\begin{figure}[t!]
\centering
\subfloat{\includegraphics[width=1\columnwidth]{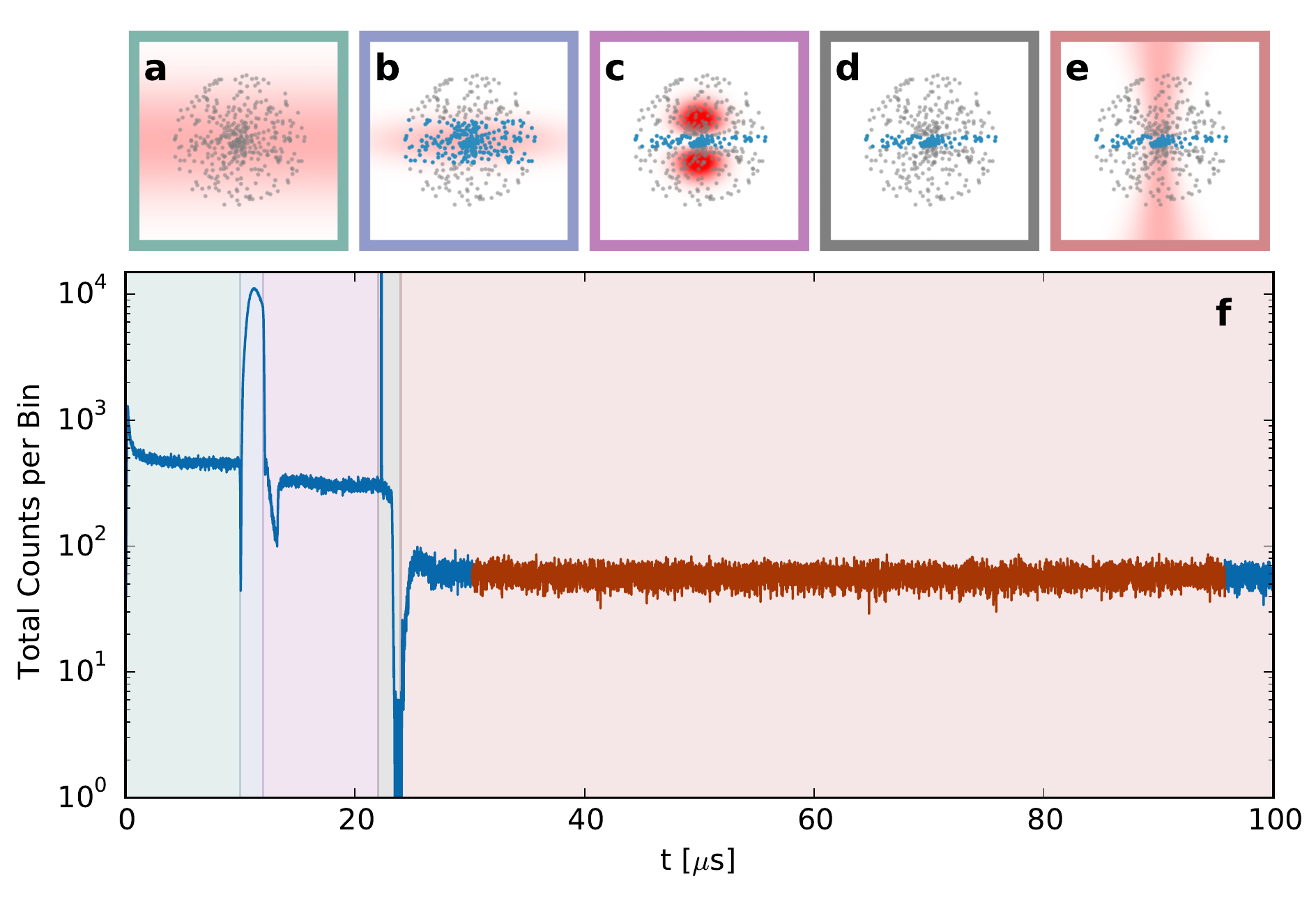}}
\caption{{\bf Slice-probe sequence beam setup and histogram of single photon events for the $g_2$ measurement. }{\bf (a-e)} shows the beam setup and the atomic state of one slice-probe cycle. The atoms color coded with blue is in $\left|F=2\right>$ ground state, and those in gray are depumped to $\left|F=1\right>$ ground state. The histogram of photons detected by the single photon counter is shown in {\bf f}. The events colored in red are used to obtain the time correlation shown in Fig. \ref{Figure:g2}. To obtain the ring down measurement, we turn on and off the probe beam 8 times during the probe stage (red background). The histogram for the modulated probe is shown in Fig.  \ref{Figure:RingdownHist}.}
\label{Figure:slicing}
\end{figure}

The atom cloud that is moved into the cavity has a dimension of $35\times 41 \times 35  \mu$m$^3$ which is much larger than the Rydberg blockade radius. The small cavity waist defines a small sample along the transverse direction to the resonator axis, but the cloud is still extended longitudinally. To confine the atom cloud outside the waist, we firstly shine a large global depump beam (Fig. \ref{Figure:slicing}a) with a waist size of 500$\mu$m. The laser is on resonance with $\left| F=2 \right> \rightarrow \left| F'=2 \right>$ transition and depumps all the atoms into $\left| F=1 \right>$ ground state. A vertical local repump beam (Fig. \ref{Figure:slicing}b) tuned to the $\left| F=1 \right> \rightarrow \left| F'=2 \right>$ transition is then switched on for 2$\mu$s.  The beam is narrow in the cavity axis in order to only repump the atoms in the center of the cloud. Finally, another slicing beam with a TEM$_{10}$ like beam profile (Fig. \ref{Figure:slicing}c) is turned on and tuned to $\left| F=2 \right> \rightarrow \left| F'=2 \right>$ transition. The node of the mode is aligned to the center of the cloud which coincides with the cavity waist. After the final slicing, the longitudinal size of the atomic cloud is reduced to $\sim$10$\mu$m. The probe and control beam are then turned on after the slicing. To avoid creating shelved Rydberg atoms, a 2 $\mu$s gap time is set between the slicing and probe process.

During the 1 ms probe time, 10 slice-probe cycles are implemented to maintain the confinement of the atomic cloud. The lattice is turned off throughout the process (primarily to avoid broadening the Rydberg level), and thus the location and size of the cloud will change due to the gravity and finite temperature. The free fall limits the total probe time to less than $1.5$ms, during which time the cloud falls by 11.5$\mu$m; within this same interval the cloud also expands to $\sim$100$\mu$m, larger than the slicing beam. As such, a global depump is performed at the beginning of each slicing sequence to remove all the atoms in the tails, as they would not be depumped effectively by the slicing beam. To maintain constant atom number in the sliced cloud over all $10$ slicing cycles in spite of cloud expansion, we repump more weakly in first cycle, and increasingly strongly over subsequent cycles. In so doing we trade-in peak density for atom number uniformity.

It bears mentioning that the magnetic field is zeroed at the cavity waist in order to minimize the dark-polariton linewidth, without necessitating optical $m_F$ pumping. The $\left| F=2 \right> \rightarrow \left| F'=2 \right>$ beams will thus optically pump some of the atoms into a dark state before depumping them to the $\left| F=1 \right>$ ground state. To break this dark-state, the global depumping beam is sent through an electro-optical modulator (EOM) with the polarization 45$^{\circ}$ off of the EOM axis; the EOM is then driven at a frequency of $2\pi\times 390$kHz, producing polarization modulation of the output light. Unfortunately this method is inapplicable to the slicing beam, because it passes through a polarizing beam splitter downstream in the optical path. To rotate the atoms out of the dark state in the slicing process, a weak polarization scrambling beam on $\left| F=2 \right> \rightarrow \left| F'=3 \right>$ transition with linear polarization is applied with the slicing beam. 

The cavity mode overlaps with the lattice beam at three points (lower waist, upper waist, and the crossing point). If a fraction of the transported atomic cloud is improperly decelerated, it can transit the crossing point or the upper waist. This sub-sample will behave as an absorbing medium, broadening the EIT feature and destroying the blockade effect. To remove these residual atoms, a ``blasting'' beam from the side which is tuned to $\left| F=2 \right> \rightarrow \left| F'=3 \right>$ is aligned between the lower waist and the crossing point. The power of this beam is set to push the atoms away before they reach the crossing point.

\section{Cavity Details\label{SI:CavityDetails}}
Our four mirror running wave resonator is arranged in a bow-tie configuration comprised of two convex mirrors and two concave mirrors in order to satisfy several opposing design constraints: To force cavity Rydberg polaritons to interact, the atomic sample size must be suitably small. The participating atomic region may be defined in two dimensions by the resonator mode cross-section, so the mode waist must be small (or radial slicing of the atomic cloud must be performed); we targeted 10-15 $\mu m$. The simplest technique for creating a stable resonator with a small waist is to use short focal length mirrors relatively close together. However, from our experience with a previous experimental cavity and its electric field filter~\cite{ningyuan2016observation}, we found that any material close to Rydberg atoms, either dielectric or metallic, will build up charges or dipoles (from Rb adsorbates), the electric fields from which unacceptably broaden Rydberg lines above $n\sim 60$. This necessitates using longer focal length mirrors so that they may be placed further apart. Then, the only way to produce a small waist is a large mode size on the surface of these mirrors. For this reason, bow-tie resonators often have a longer upper arm with relatively flat mirrors so that the beam may expand due to diffraction, thereby increasing the beam size at the lower mirrors and so reducing the waist size. Having a longer upper arm, however, was unacceptable for two reasons: First, the resonator linewidth would decrease (at constant finesse), which proportionally decreases the data collection rate, (and the autocorrelation-data rate goes as the square of this rate). Second, the resonator is loaded into our vacuum chamber through a 62 mm diameter tube, which therefore sets an absolute maximum exterior size. Both of these limitations could be avoided, however, by utilizing convex mirrors in the upper arm; the defocus that they create acts, for our purposes, equivalently to diffractive expansion. 

\begin{figure}[t!]
\centering
\subfloat[]{
	\includegraphics[width=0.5\columnwidth]{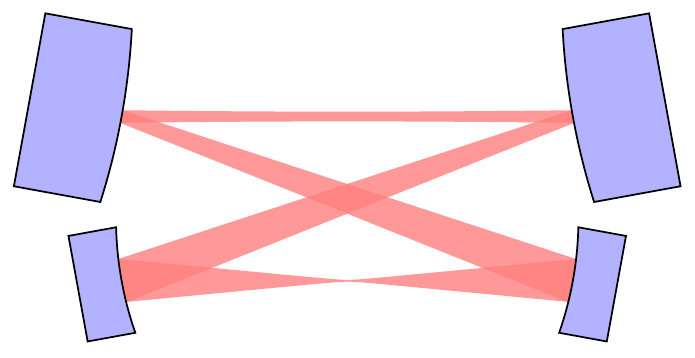}
    \label{Figure:CavDetails_Vis}
    }\,
\subfloat[]{
	\includegraphics[width=0.35\columnwidth]{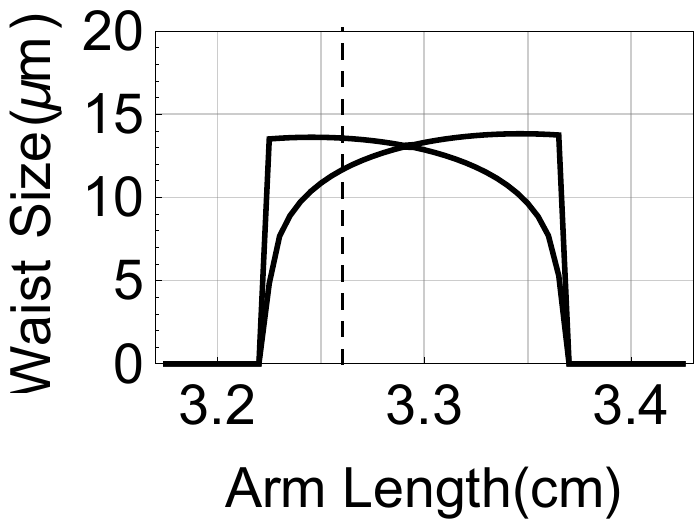}
    \label{Figure:CavDetails_StabilityRegion}
    }\\
\subfloat[]{
	\includegraphics[width=1\columnwidth]{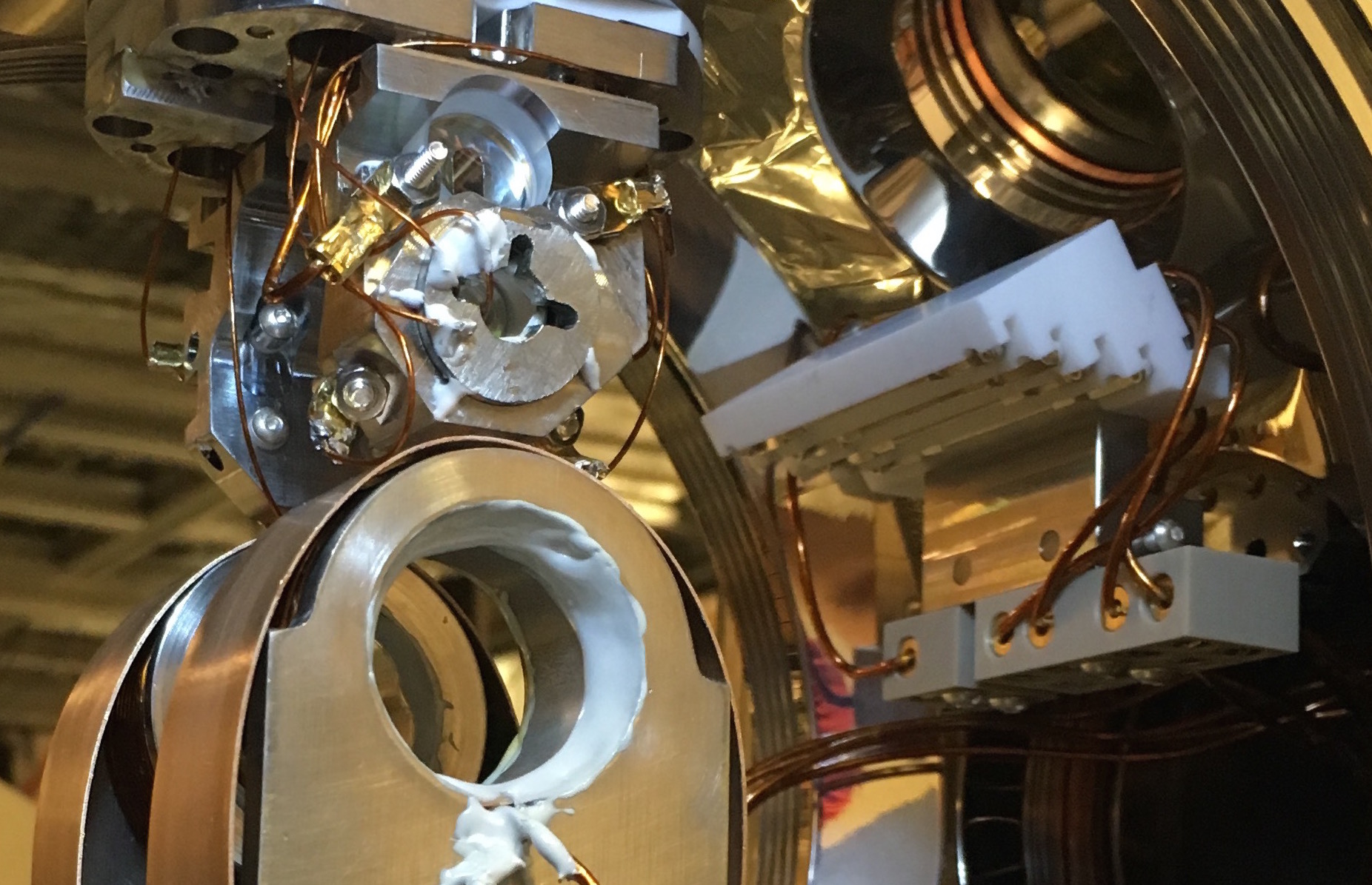}
    \label{Figure:CavDetails_Photo}
    }
\caption{(a) A schematic of the resonator. The transverse size of the mode reflects the results of numerical calculation given the resonator geometry and is represented to scale but for a factor of 10 magnification compared to the mirror dimensions and locations. (b) The numerically calculated mode waist sizes are plotted as the resonator length is increased with constant aspect ratio. The cavity is stable over a 1.5 mm range, although near the edges of that range, it becomes significantly astigmatic. We set out to build the resonator at the middle where the modes waists are equal; however, assembly imperfections resulted in a somewhat shorter actual length, indicated by the dashed line. (c) A photograph of the in-vacuum experimental apparatus showing the resonator (top), the MOT coils (bottom), and Rubidium dispensers in their macor mount (right). The lower resonator mirror is mounted on a piezo tube which is glued to the lower circular steel disk on the resonator structure. Through the upper mirror, a narrow slot may be seen. Such slots are the only opening to the front of the mirrors and serve as a passive electric field filter to reduce the effect of charges that may build up on the mirrors' dielectric surfaces. Six out of the eight electrodes are visible: four from the outside showing their attachment to wires via gold plated crimps, and two on the inner surface of the cavity. The heads of the screws form the surface of the electrode.}
\end{figure}

\begin{figure}[h]
\includegraphics[width=0.8\columnwidth]{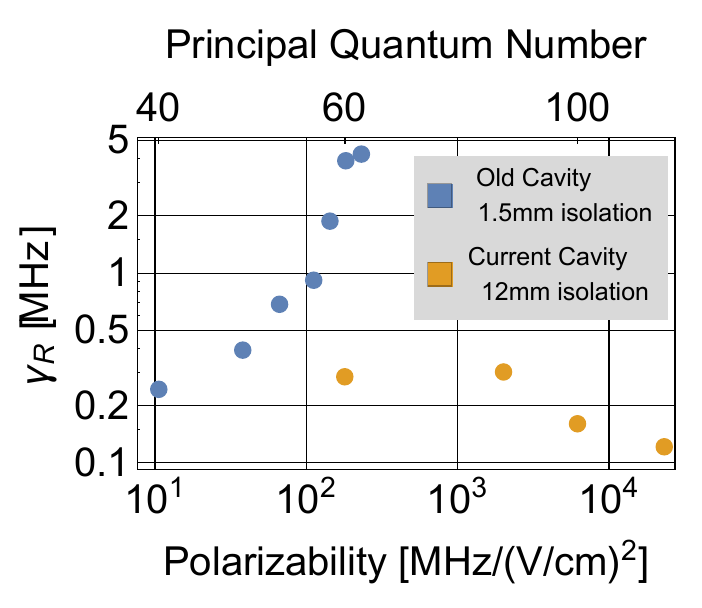}
\caption{The previous experimental resonator had material $\sim$1.5mm away from the atomic sample at the resonator waist. Even with eight electrodes to control the electric field environment, this led to significant electric fields and gradients at the atoms. These fields increase the Rydberg state loss, $\gamma_R$, as seen in fits to broad EIT spectra (blue). The current experimental resonator places the nearest material over 12 mm away from the atom, with most material well over 1cm away. After optimizing the electric field environment via electrodes, we find that the Rydberg state loss decreases slightly with principal quantum number.}
\label{fig:gammar_vs_pol}
\end{figure}

With numerical modeling, we arrived at an acceptable configuration of the resonator given these constraints, providing a $\sim 13 \mu m$ waist at least 12 mm away from the nearest surface. To provide passive electric field attenuation, the steel mounting structure fully encloses the locking piezo, which can be driven up to 1 kV, and the front surface of every mirror is covered except for small aperture at the mode location. Additionally, eight screw-head electrodes provide active electric field control, and the two separate halves of the mounting structure may each be set to an arbitrary electric potential. In total, this provides ten degrees of freedom to control the eight independent electric field and electric field gradient components. Finite element analysis based on 3D CAD designs then provide the conversion matrix between applied electrostatic potentials and the electric field and gradients at the mode waist. Since the number of electrodes exceeds the number of controlled components, we calculate and use the `optimal' conversion matrix which minimizes the applied electric potentials. 

Figure \ref{fig:gammar_vs_pol} shows the evolution of Rydberg linewidth with principal quantum number over two resonators: one with material $\sim1.5$mm~\cite{ningyuan2016observation} from the laser-cooled atomic sample, and the optimized structure employed in this work (see Fig. \ref{Figure:CavDetails_Photo}), where the closest surface is 12 mm from the atomic sample. This change produces a $\sim3000$-fold reduction in field-induced broadening, enabling us to enter the strongly-interacting regime.

The resonator's upper mirrors are plano-convex with a 50 mm radius of curvature while the lower mirrors are concave with a 25 mm radius of curvature. The two upper mirrors and the non-piezo lower mirror have a custom coating provided by Layertec GmbH, specified to have a 99.9\% reflectivity at both 780 nm and 1560 nm, while having $>$95\% transmission at 480 nm. The other lower mirror has a coating by Advanced Thin Films with much higher reflectivity of 99.995\%. While the optimal finesse for this resonator would be $\sim 2100$, contamination during resonator alignment resulted in additional loss and a cavity finesse of $F=1480(50)$. The free spectral range is 2204.6 MHz, measured with an EOM sideband, and the polarization eigenmodes are approximately linear and split by 3.6 MHz. The measured absolute on-resonance transmission, in-coupling through one upper mirror and out-coupling through the other upper mirror, is $T_{cav}=8.2$\%, providing an estimate of the single pass transmission through an upper mirror $T_{upper}=\sqrt{T_{cav}}\frac{\pi}{F}=0.061$\%, and a probability of out-coupling any given intra-cavity photon through a top mirror of $P_{outcoupling}=\frac{F T_{upper}}{2\pi}=14$\%.The detection path efficiency (after resonator out-coupling, and excluding the detector) of 92\% is dominated by narrow line filters used to block external background light and down-converted 480 nm photons (see SI ~\ref{SI:spdc}). The output is split on a 50:50 beam-splitter, and the resulting beams are aligned (with 97\% efficiency) to two single photon counting modules with a quantum efficiency of 55\%. The total detection path quantum efficiency (for both detectors together) is thus $\eta_{tot}=7.3$\%. From this, we calculate a conversion between photon detection rate and intra-cavity photon number of $708\times 10^3 s^{-1}$ $/$photon.


\section{Cavity Locking\label{SI:CavLocking}}
We lock the experimental resonator to an arbitrary detuning from the atomic resonant frequency, first by locking a frequency doubled 1560 nm laser to an ultra stable notched Zerodur cavity, (Stable Laser System Model VH6020-4 with linewidths of $2\pi\times 55$ and $2\pi\times 41$ kHz at 780nm and 960nm nm, respectively), and then lock the experimental resonator to the 1560 nm laser. The 780 nm lock to the ultra-stable cavity is accomplished via a two tone Pound-Drever-Hall technique. An EOM modulates the 780 nm light at two frequencies, one at 10 MHz, the other an arbitrary DDS-generated, computer controlled frequency between 50 MHz and 1.5 GHz. Demodulation of the reflection signal at 10 MHz provides three locking features, one at the cavity resonant frequency, one at the cavity resonant frequency plus the DDS frequency, and one at the cavity resonant frequency minus the DDS frequency. By locking to either the upper or lower sideband locking feature, this scheme allows locking the 780 nm probe (and hence 1560 nm) carrier frequency to an arbitrary frequency relative to the atomic resonance. 

The experimental resonator is locked to the 1560 nm laser via two tone frequency modulation spectroscopy of the resonator, with feedback controlling the length of a piezo tube on which one of the lower resonator mirrors is mounted. The error signal generation is similar to the Pound-Drever-Hall technique used with the ultra-stable cavity, except that we demodulate the transmission signal rather than the reflection because the reflection is not directed through a window of the vacuum chamber. While this in principal lowers the response of the lock above the resonator's 1560 nm linewidth (9MHz), the locking bandwidth is in practice limited much sooner by piezo-mount resonances at few kHz level. This permits locking the resonator to an arbitrary detuning from the atomic resonance.


Slow drifts in temperature cause the resonator piezo locking voltage to drift, with a single $1560$ nm free spectral range corresponding to $\sim$500 V on the piezo (the piezo is non-linear, so this depends on where in the $1$ kV range of the piezo the FSR is measured). In particular, daily modulation of the Rubidium dispenser current and variations in the MOT coil current cause heating of the resonator. Because the resonator is thermally well isolated from the environment, the thermal relaxation time constant is $\sim$1 hour. The thermal impulse is also significant: when the dispensers are turned on, the resonator drifts through 4-5 $1560$ nm free spectral ranges before settling. A particularly concerning effect from this voltage drift is the variable electric field experienced by Rydbergs at the mode waist. We expended considerable effort to isolate the atoms from charge buildup and the piezo voltage, but still found that, at 100S, a drift of $\sim$30 V on the piezo would shift and broaden the dark polariton significantly. At 121S, only a few volts of piezo drift are necessary to shift out of EIT resonance.

In order to remove this instability, we implement slow digital feedback on the piezo locking voltage by heating the steel resonator structure using a pulse-width modulated 980 nm laser with $\sim$1 W of power focused to a $\sim$1 mm diameter spot. This slow feedback stabilizes the locked piezo-voltage to a set-point of 60 V, with RMS error of $<$0.05 V.


\begin{figure}
\includegraphics[width = .8\columnwidth]{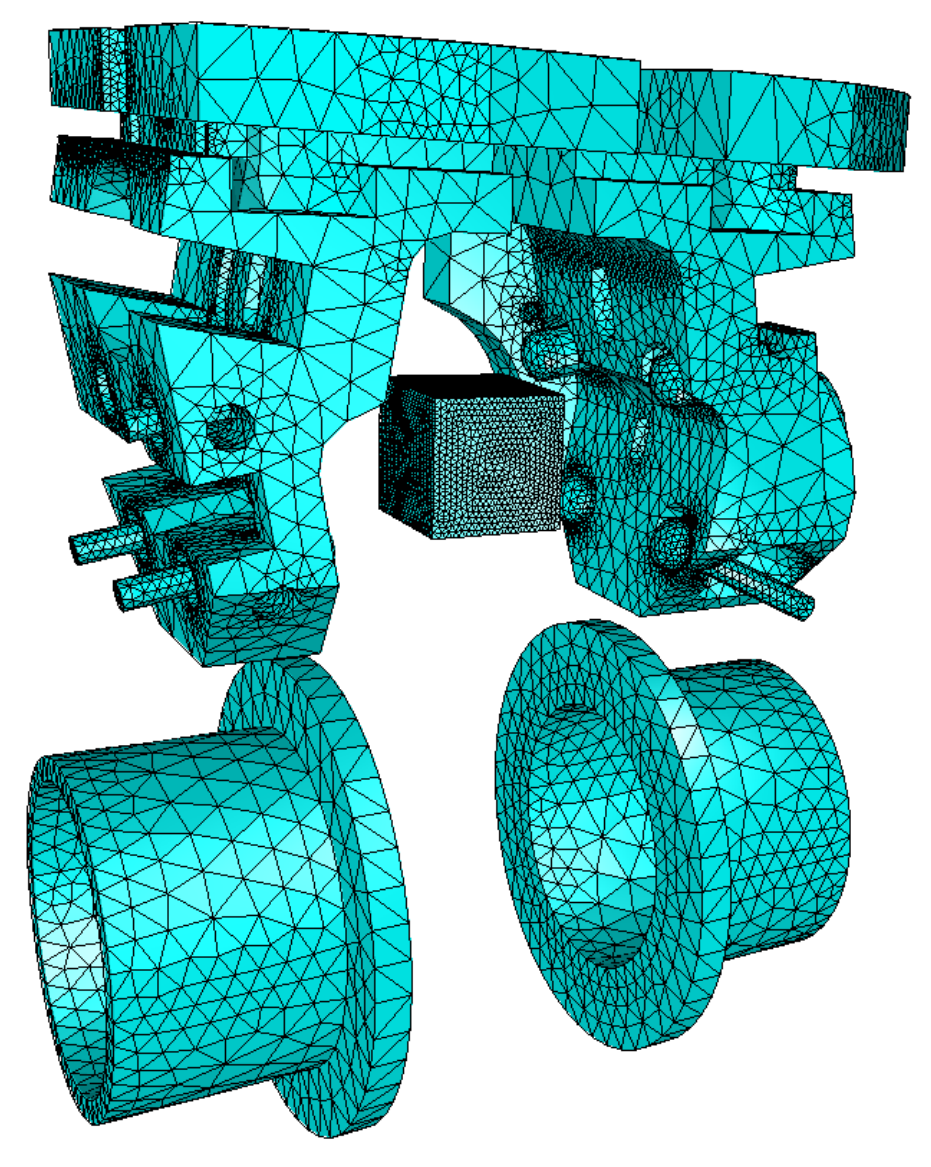}
\caption{A 3D CAD .stl model generated in SolidWorks is imported into ElmerGUI finite element analysis package via the Salome mesh generation software. The steel mounting structure, eight electrodes with their macor spacers, high voltage piezo, and MOT coils are included in the model; an additional fictitious cube with a finer mesh is specified near the atom location in order to increase the precision of the result. Each insulator and conductor is assigned a permittivity or voltage boundary condition respectively, with a large bounding sphere enclosing the entire model (not shown) given an `effective infinite' boundary condition. The electric field at the atom position is calculated with each metal body set to 0V except for one set to 1V, and a python script then iterates over which metal body is set to 1V. The results provide a matrix converting electrode potentials to electric field at the atoms. This may then be inverted and used in the experimental control apparatus to convert a desired electric field (or gradient) into applied voltages.}
\end{figure}

\section{Fluorescence/Parametric down-conversion background of 480nm $\rightarrow$ 780nm photons\label{SI:spdc}}
We employ free-space single photon counting modules (SPCMs) to improve our detection path efficiency. Careful elimination of environmental backgrounds enables us to reach the manufacturer-specified 50 Hz dark count rate of the detectors. However, in the presence of the 480nm control field, we experience 200 Hz of additional counts, even with interference filters (blocking 300-1200nm, save a $\pm \sim 1$ nm interval around 780 nm) placed in the 480nm path immediately before the vacuum chamber. Removing this background required \emph{ultra-narrow} filtering around 780nm after the vacuum chamber, but before the SPCMs, which we achieved with two 785 nm clean-up filters (each $\pm \sim 1$ nm wide, with sub-nm edges) in the detection path, and appropriately tilted to narrow the filtering bandwidth by an order of magnitude, leaving an excess background of only 20 counts/second.

The source of this background is either fluorescent or spontaneous parametric down-conversion of 480 nm photons \emph{within the resonator mirrors and vacuum chamber windows} to 780 nm; this effect is not observable with single-mode fibers in the detection path, as both fluorescence and SPDC are highly multi-mode, and is thus filtered out. The down-conversion almost certainly occurs within the resonator mirrors, because the control beam is focused to a tight spot at the resonator waist, the intensity of the beam in the resonator mirrors is much higher than in any other piece of glass.

\section{Inhomogeneous broadening and Performance Limitations\label{SI:inhomog}}

We observe a slight disagreement between the measured $T_1=590$ ns (and corresponding linewidth of $\gamma_{EIT}=2\pi \times 270$ kHz) and the width of the EIT feature in Fig. ~\ref{Figure:SetupFigure}b of $\gamma_{EIT}=2\pi \times 400$ kHz, indicating $\sim 2\pi\times 130$ kHz of inhomogeneous broadening. This inhomogeneous broadening impacts the EIT feature by increasing its linewidth and suppressing its height, but does not substantially impact the bright polariton features due to their reduced Rydberg admixture and larger intrinsic linewidth from P-state-admixture.

\begin{figure}[t!]
\centering
\subfloat[]{
	\includegraphics[width=0.5\columnwidth]{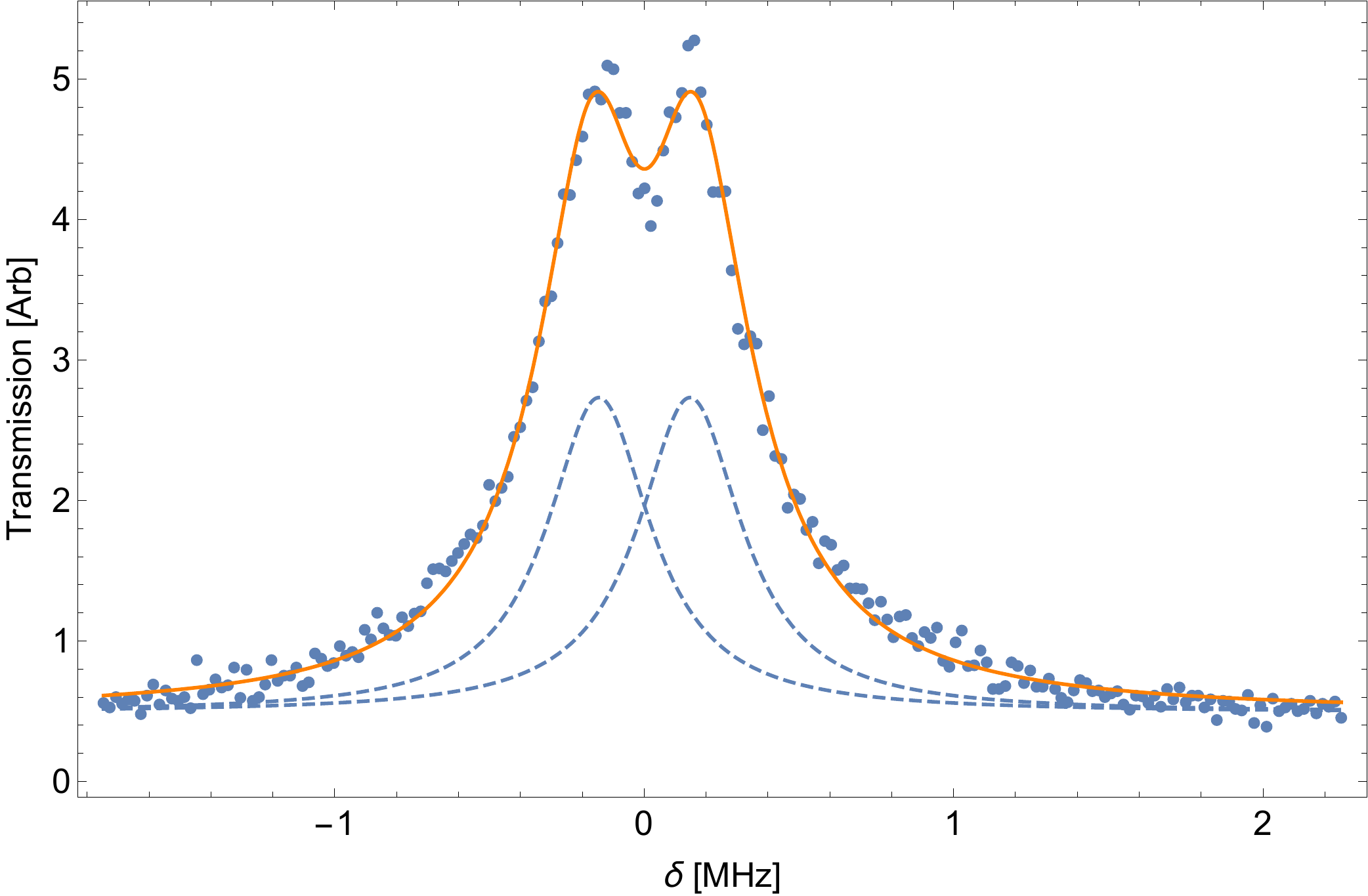}
    \label{Figure:DoublePeakEIT}
    }
\caption{\textbf{Structure of the EIT transmission peak.} The resonator transmission is plotted vs. probe detuning, on the $100S_{\frac{1}{2}}$ EIT feature, at zero E- and B-field. Observed splitting of the feature, of magnitude $2\pi\times 294$ kHz, likely results from weak Zeeman-coupling to a nearby degenerate Rydberg $m_J$ state.}
\end{figure}

Indeed, a more careful examination of the central EIT feature, shown in Fig. ~\ref{Figure:DoublePeakEIT}, reveals a small splitting. We postulate that this arises from a weak admixture of a nearly-degenerate Rydberg state, potentially by a weak Zeeman field, or Hyperfine coupling. Another possibility is a near-degenerate ultra-long-range Rydberg molecular state, explored recently by the Hofferberth group~\cite{mirgorodskiy2017electromagnetically} at slightly lower principal quantum numbers.

\section{Calculation of Dark Polariton Rabi Oscillation frequency}
\label{SI:DSPrabifreq}
\begin{figure}
\includegraphics[width=.7\columnwidth]{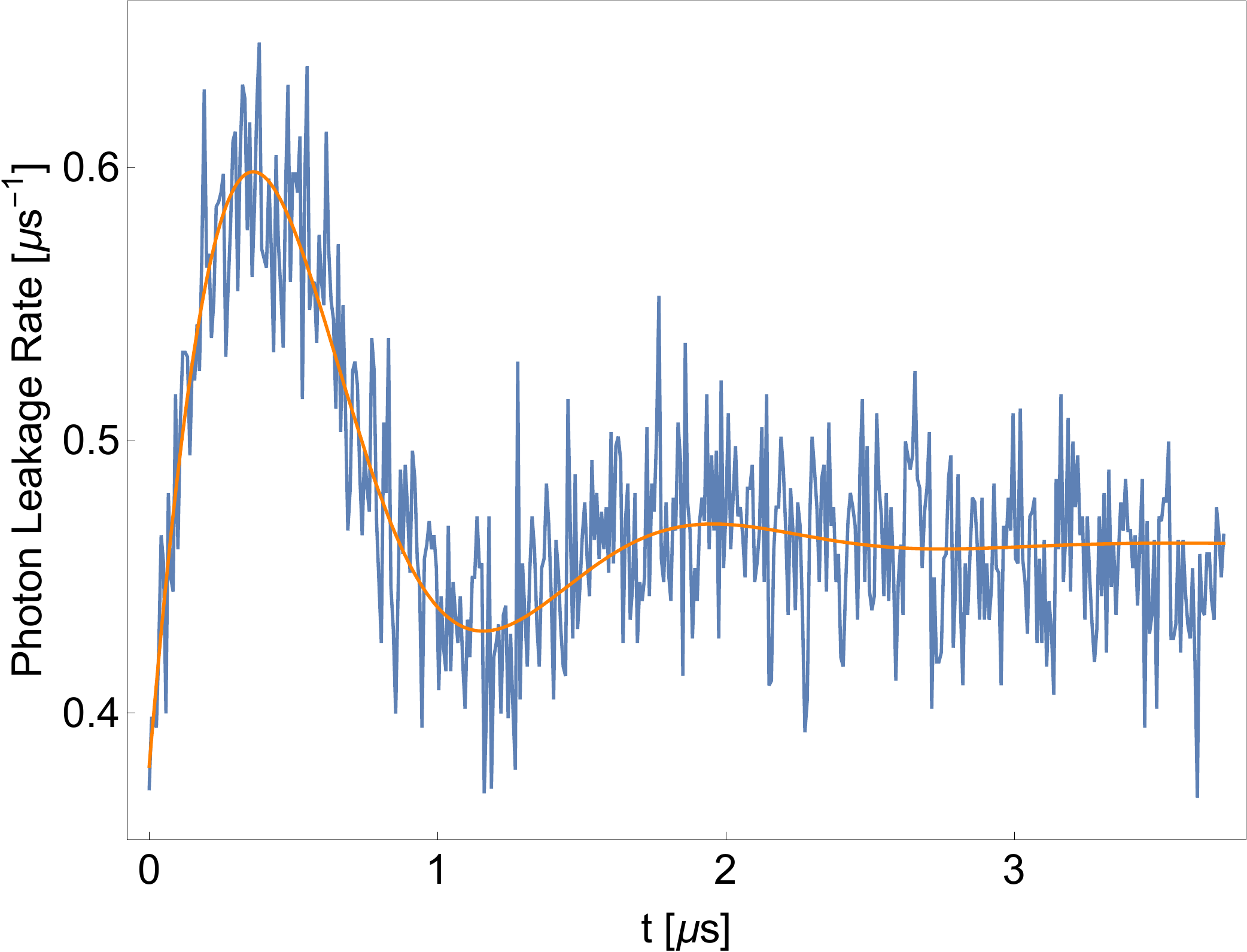}
\caption{\textbf{Polariton Rabi Oscillation and Optical Bloch Equation Fit.} The Rabi oscillation between 0 and 1 dark polaritons (Fig. \ref{Figure:ringdown}a, highest power) is plotted subsequent to the bright polariton dynamics of the first $\sim 440$ns, along with a fit to the Optical Bloch equation.}
\end{figure}
To extract the measured Rabi oscillation frequency between zero and one polaritons from the highest probe power data in figure ~\ref{Figure:ringdown}b, we fit the ring-up curve with the solution of the optical Bloch equation starting with $\rho(t=0)=|0\rangle\langle 0|$:
\begin{equation}
\rho_{ee}(t)=\frac{1}{2}-\frac{\gamma^2}{\gamma^2+8\omega^2}\left[1+\frac{4\omega^2}{\gamma^2}e^{-3\gamma t}\left(\cos{\chi t}+\frac{3\gamma}{\chi}\sin{\chi t}\right)\right]\\
\end{equation}

Here $\chi\equiv 2\sqrt{\omega^2-\gamma^2/4}$; the fit to the data yields $\omega = 2\pi \times 318$ kHz, corresponding to the observed $2\pi$-pulse time of approximately $1.2\mu$s, corrected for the rapid decay $\gamma=2\pi\times 390$ kHz.

This $\omega$ should be compared to the Rabi frequency $\Omega_p$ predicted based on the cavity driving, which we extract from the incident photon rate, $R_{inc}$:
\begin{equation}
\Omega_p = \sqrt{\frac{R_{inc}}{4}\kappa \cos^2(\theta)}.
\end{equation}
At the highest power, $R_{inc}=$ 22 MHz, yielding $\Omega_p= 2\pi\times 400(40)$ kHz, within $\sim 2\sigma$ of the measured value above.

\section{Calculation of high power steady state dark polariton number}
\label{SI:DSPnumber}

The intracavity dark polariton number is calculated as:
\begin{equation}
n_D = \frac{R_D}{\kappa \cos^2(\theta)}
\end{equation} 
where $R_D = 0.18$ $\mu$s$^{-1}$ is the steady state rate of photons emitted from dark polaritons obtained from Fig. 4(c), solid black curve, at the highest probe power, $\kappa = 2\pi\times 1.55$ MHz is the cavity linewidth, and $\cos^2(\theta) = \frac{\Omega^2}{g^2+\Omega^2}$ is photonic fraction of the dark polariton, with $g= 2 \pi \times 5.5$ MHz and $\Omega = 2 \pi \times 2.0$ MHz. This yields $n_d = 0.16(8)$.

\begin{figure}
\subfloat{
	\includegraphics[width=1\columnwidth]{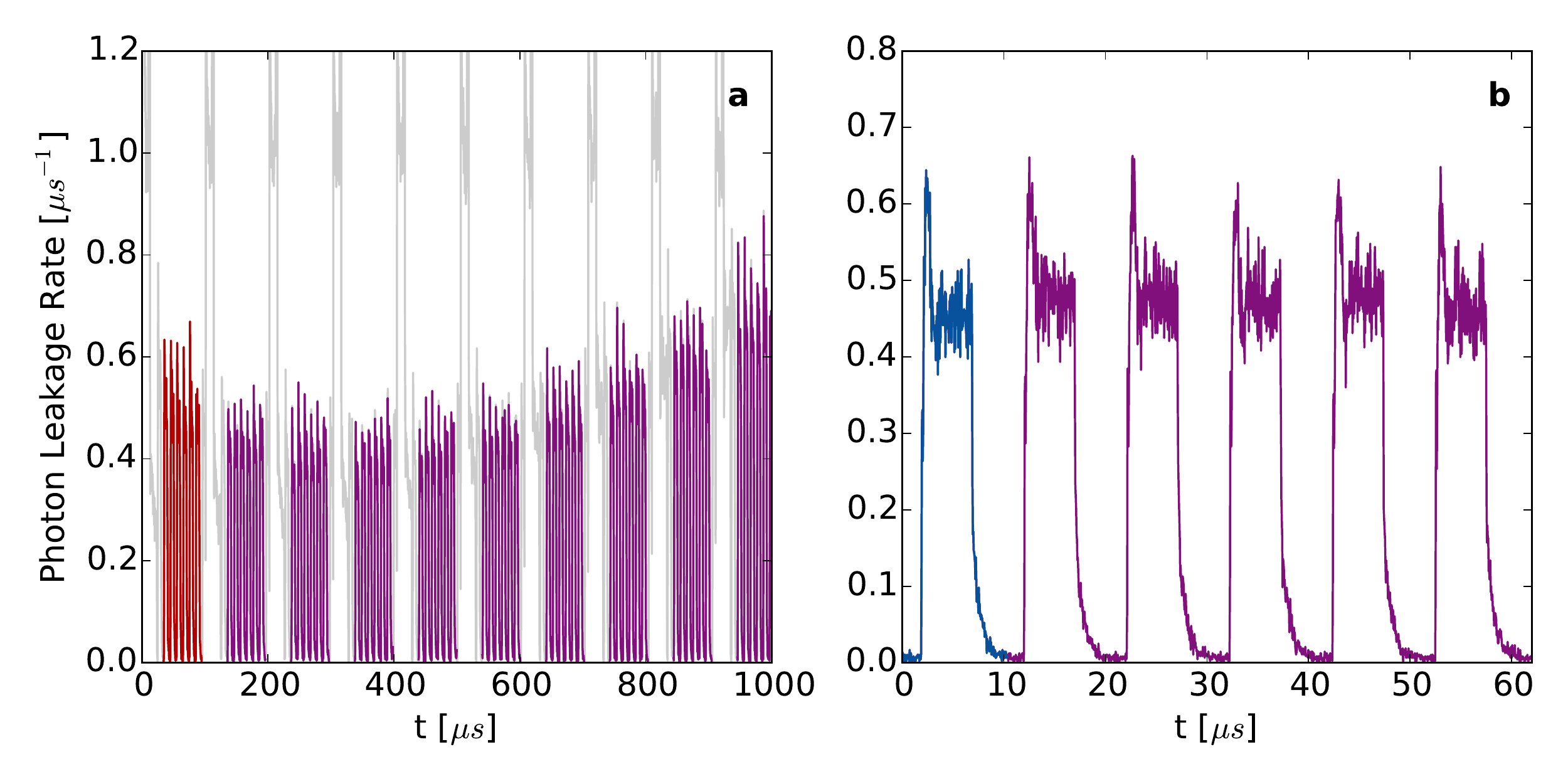}
    }\\
\subfloat{
	\includegraphics[width=1\columnwidth]{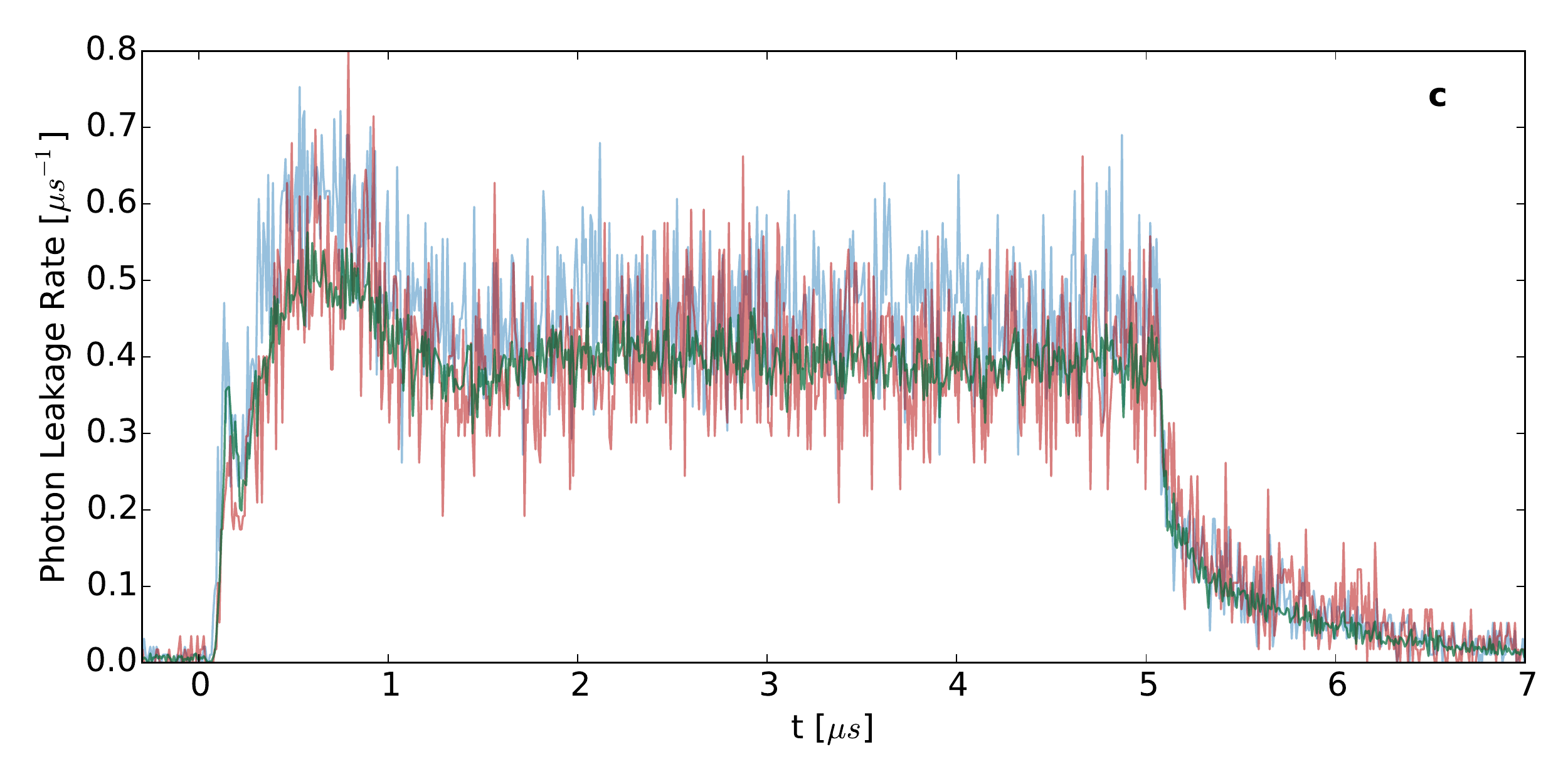}
	}
\caption{\label{Figure:ringdown_hist} \textbf{Ringdown Count Arrival Histogram. (a)} The entirely unfolded ringdown histogram of photon arrival times (data from Fig. \ref{Figure:ringdown}a, highest power) is built up over 16,000 individual runs of the experiment. Ten `reslicing' intervals are apparent, with the data used highlighted in purple, and the first reslicing interval in orange. The smooth and relatively minor fluctuation in the transmission rate over the course of the 1 ms total probe time indicates successful reslicing with constant atom number and the lack of shelved Rydbergs which would reduce the transmission. \textbf{(b)} The same data are folded over each reslicing interval, with the excess depump and repump photons as well as the first and last probe cycles cut out. This leaves six 10 $\mu$s probe cycles, the first of which is highlighted in blue. When all six are folded on top of each other, they yield data for Fig. \ref{Figure:ringdown}. The steady transmission is constant over these cycles indicating that Rydbergs are not being shelved. \textbf{(c)} To investigate the possible buildup of stray Rydberg atoms, we plot the folded data of the first reslicing interval in orange, the folded data of the first probe cycle from all reslicing intervals in blue, the entire folded dataset (same as Fig. \ref{Figure:ringdown}a, highest power) in green. If stray Rydbergs were developing during each 1 ms run or within each reslicing interval, either the orange or blue curves should saturate to a significantly higher leakage rate.}
\label{Figure:RingdownHist}
\end{figure}

Naively, we would expect $n_D = 0.5$, arising from an equilibrated, strongly driven two-level system in the presence of loss. That the dark polariton number saturates significantly lower than this is most likely due to a population of Rydbergs that is entirely decoupled from the resonator:

In Fig. ~\ref{Figure:AdditonalMasterEqnFit}, we plot the results of a master equation fit to the observed data in Fig. ~\ref{Figure:ringdown}a of the main text, for all drive powers. We employ shelved Rydberg probabilities of $\alpha_{ryd}=\{0.42,0.5,0.5,0.42\}$, and $\gamma_{ryd}=2\pi\times 120$kHz; the good agreement between theory and experiment validates this model. A probable source of this background is rotation of the collective Rydberg excitations into a collective Rydberg state decoupled from the cavity (either due change of collective symmetry, Zeeman shifts, Doppler, etc.). Another possibility more technical in nature: leakage of the $\sim 780$nm slicing fields during the probing process, creating a background of Rydbergs. All of these processes result in a Rydberg excitation which cannot be optically detected, but which nonetheless precludes injection of a further dark polariton.

The suppression of $n_D$ below $0.5$ is also potentially attributable to several other less-likely scenarios:
\begin{figure}[b]
\includegraphics[width = .8\columnwidth]{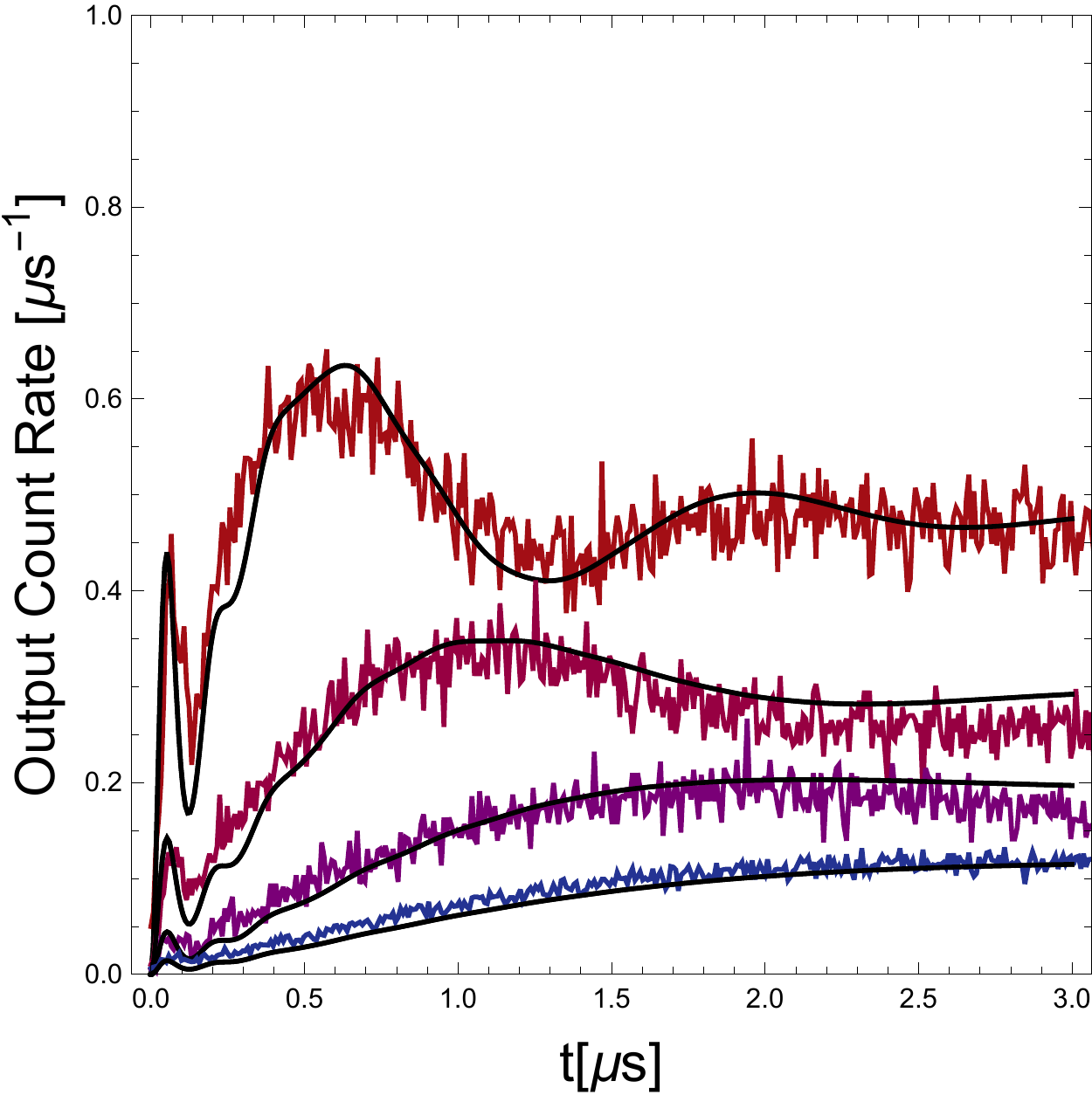}
\caption{\label{Figure:AdditonalMasterEqnFit} \textbf{Master Equation Model to Ring-Up Data} A master equation model of the Ring-up dynamics at all probe powers reflecting a shelved Rydberg which blockades dark-polariton excitation with a probability of $\sim 50\%$.} 
\end{figure}
\begin{enumerate}
\item Long-lived Rydberg atoms are created by breaking dark polaritons (in, for example Rydberg-Rydberg collisions). These ``shelved'' Rydbergs would blockade the sample, resulting in a lower saturated dark polariton number; to be long-lived, they would have to reside in a different Rydberg level which is decoupled from the blue field. We check and discredit this possibility based on the photon arrival histograms for the high power ring-down data shown in Fig. \ref{Figure:ringdown_hist}. There is no apparent reduction in resonator transmission (the tell-tale sign of Rydberg-buildup) within a single ring-up/ring-down cycle, over the 6 cycles that comprise a slicing interval, or over the 10 slicing intervals that comprise a full experimental cycle.
\item Multiple dark polariton modes contribute to the (total) saturated polariton number of $n_D = 0.5$, but we are only able to detect one of these modes. For example, the dark polariton of the backwards-propagating cavity mode is degenerate with the forward-propagating dark polariton we excite and detect. The strong interactions between polaritons prohibit a second polariton from entering the cavity; however, it may do so virtually, and thereby induce coupling between the single excitations in the forward- and backwards- dark-polariton manifolds through the (interaction shifted and broadened) two-polariton manifold. The small Rydberg admixture of a bright polariton could similarly induce such a forward-to-backwards conversion of a dark polariton. Reasonable values for an interaction shift along with the known dark polariton Rabi frequency provide a rough estimate of the equilibration time between two degenerate dark polaritons as short as $\sim 10$ ns, while the observed curves place an upper bound on such an equilibration time of $\sim 100$ ns.
\item Inhomogeneous broadening causes jitter such that the probe is significantly detuned from the dark polariton resonance some of the time. This artificially reduces the detected rate of photons but does not change timescales such as the dark-polariton Rabi-oscillation frequency or dark-polariton lifetime.
\end{enumerate}

\begin{widetext}
\section{A Simple Master-Equation Effective Model of Cavity Rydberg Polaritons}
\label{SI:SimpleMasterEqn}
Here we present the simple master-equation model that we employ to model the dynamics of the cavity ring-up and ring-down. Instead of employing the full model of $N$ 3-level atoms coupled inhomogeneously to both the resonator and control field, and interacting with one another in a position-dependent manner (as in the later SI section), here we simply assume all atoms are identical, with an interaction in the Rydberg state of strength $U$. Accordingly, the Hamiltonian is:

\begin{align}
H=\omega_c a^\dagger a+\omega_e p^\dagger p+\omega_r r^\dagger r+g\sqrt{N}(a^\dagger p+p^\dagger a)+\Omega (p^\dagger r+r^\dagger p)+\Omega_D(a^\dagger+a)+\frac{U}{2}r^\dagger r^\dagger r r
\end{align}

The evolution of the reduced density matrix $\rho$ is given by a Lindblad master equation:
\begin{align}
\dot{\rho}        &= -i [H,\rho]+\frac{\kappa}{2}\mathcal{L}(\rho,a)+\frac{\Gamma}{2}\mathcal{L}(\rho,p)+\frac{\gamma_r}{2}\mathcal{L}(\rho,r)\\
\mathcal{L}(\beta) &\equiv \beta\rho\beta^\dagger-\frac{1}{2}\{\rho,\beta^\dagger\beta\}
\end{align}

The bosonic operators: $a^\dagger$($a$) creates (destroys) a resonator photon; $p^\dagger$($p$) creates (destroys) a P-state atom; and $r^\dagger$($r$) creates (destroys) a Rydberg atom. Our numerics begin at $t=0$ with initial condition $\rho(t=0)=|0\rangle\langle 0|$.

Here $\kappa = 2\pi\times 1.55$ MHz is the cavity linewidth, $\Gamma = 2\pi \times 6$ MHz is the atomic P-state linewidth, $\gamma_r = 2\pi \times 200$ kHz is the collective Rydberg-state linewidth~\cite{ningyuan2016observation}, $\omega_c = 2 \pi \times 0$ kHz is the resonator frequency in the frame rotating with the probe beam, $\omega_e = 2 \pi \times 0$ kHz is the energy of the atomic P-state in the frame of the probe beam, and $\omega_r = 2\pi \times 0$ kHz is the atomic Rydberg state energy in the frame of the probe and control beams. $g\sqrt{N} = 2 \pi \times 4.3$ MHz is the collective light-matter coupling, $\Omega = 2 \pi \times 1.5$ MHz is the Rabi frequency of the control field, $U = 2\pi \times 4$ MHz is a parametrization of the typical Rydberg-Rydberg interaction strength, and $\Omega_D = 2 \pi \times 1.3$ MHz is the strength of the resonator driving field. 

The atom-photon coupling and the control Rabi frequency  both fit to significantly lower values than fit to the transmission spectrum in Fig. ~\ref{Figure:SetupFigure}b indicates. This is directly related to the saturation of the dark polariton number below $n_D=0.5$, which this model requires. By reducing $g\sqrt{N}$ and $\Omega$, the model increases the bright polariton population to match the early peak in the lineshape characterized by artificially low saturated dark polariton number, while the free overall scaling then allows the fit to match the total output. This model, then, is useful for phenomenologically identifying features associated with bright polaritons versus dark polaritons and to understand the parametric dependence of the system performance, e.g. atom number or probe power, but should not at present be trusted quantitatively.

A slightly more sophisticated model allows for a background of ``shelved'' Rybergs (really only one Rydberg, at most) that does not dynamically evolve, but which may nonetheless blockade the system. Implementing such a model with shelved Rydberg probabilities ranging from $\alpha_{ryd}=0.42...0.5$, and $\gamma_{r}=2\pi\times 120$kHz, enables us to fix all other experimental parameters to those observed in Fig. ~\ref{Figure:SetupFigure}b: $g\sqrt{N}=2\pi\times 5.5 MHz$ and $\Omega=2\pi\times 2.0 MHz$, plus the cavity-drive Rabi frequency calculated from the bare resonator transmission. This model reproduces the observed steady state dark polariton emission rate at all drive powers, and agrees qualitatively with all features in Fig. ~\ref{Figure:AdditonalMasterEqnFit}.

To implement this model, we must add another Rydberg mode $b$ corresponding to the shelved Rydbergs, with no decay. The Hamiltonian is then given by:
\begin{align}
H&=\omega_c a^\dagger a+\omega_e p^\dagger p+\omega_r n_{ryd}+g\sqrt{N}(a^\dagger p+p^\dagger a)+\Omega (p^\dagger r+r^\dagger p)+\Omega_D(a^\dagger+a)+\frac{U}{2}(n_{ryd}^2-n_{ryd})\\
n_{ryd}&=r^\dagger r + b^\dagger b
\end{align}

We evolve our dynamics with an initial state: $\rho(t=0)=(1-\alpha_{ryd}) |0\rangle\langle 0|+\alpha_{ryd} b^\dagger|0\rangle\langle 0|b$.

\section{An Efficient Approach to Calculate $g_2$ and Linear Transmission\label{SI:NumericalMethods}}
Here we provide an efficient way to calculate the transmission, up to first non-vanishing nonlinear order, of a resonator coupled to interacting atoms. The linear part of the transmission requires no matrix inversions, while the first nonlinear correction (the intensity-intensity correlator $g_2$) requires a linear solve of a matrix of dimension $\sim N\times N$, where $N$ is the number of atoms, much better than the n\"aive expectation of an $\sim N^2\times N^2$ matrix. In particular, we demonstrate that the sparsity of the (a) singly- and (b) doubly- excited manifolds may be harnessed \emph{explicitly} to completely remove the need for a matrix inversion, or vastly reduce the complexity of the necessary inversion, respectively.

We first explore the closed-form calculation of the transmission, demonstrating not only that a matrix inversion is unnecessary, but also that the transmission may \emph{still} be written in closed form in the presence of a finite sample size and probe beam size. This first computation is relatively straightforward, and sets the stage for the messier $g_2$ calculation in the two-excitation manifold. We then explore the reduction of the $g_2$ calculation from a $(2N)^2 \times (2N)^2$ sparse-matrix inversion to a $(2N) \times (2N)$ dense-matrix inversion.

\subsection{Non-Hermitian Perturbation Theory}
Starting from a vacuum state $|0\rangle$ with no photons in the cavity, and each atoms in its ground state $G$, the system may be driven via a cavity probe field into a steady state given by $|\psi\rangle=[\hat{H}-\hat{n}\omega]^{-1}\hat{V}[\hat{H}-\hat{n}\omega]^{-1}\hat{V}|0\rangle$, where $\hat{n}$ is the operator that determines how many excitations are in the system, $\omega$ is the frequency of the incident cavity probe, and $\hat{V}=\Omega_p(\hat{a}^\dagger+\hat{a})$, where $\Omega_p$ parameterizes the strength of the probe field. In what follows the cavity propagation axis is $\hat{z}$.

The Hamiltonian is given by:
\begin{eqnarray}
\hat{H}=\omega_c \hat{a}^\dagger \hat{a}&+&\sum_m \left[\omega_e \hat{\sigma}^{ee}_m+\omega_r \hat{\sigma}^{rr}_m+\left(g_m \hat{a} \hat{\sigma}^{eg}_m+h.c.\right)+\left(\Omega_m\hat{\sigma}^{re}_m+h.c.\right)\right]\\
&+&\frac{1}{2}\sum_{m\neq n}\hat{\sigma}^{rr}_m \hat{\sigma}^{rr}_n U(|\vec{x}_m-\vec{x}_n|)\nonumber
\end{eqnarray}

Here $g_m=g e^{-|\vec{x}_m\times\hat{z}|^2/w_c^2}$ is the vacuum-Rabi coupling strength of atom $m$ to the resonator mode of waist $w_c$; $\Omega_m=\Omega e^{-|\vec{x_m}\times\hat{z}|^2/w_b^2}$ is the strength of the laser-induced coupling between atomic states $e$ and $r$, due to a (P-to-Rydberg control) beam of waist $w_b$. $\hat{a}^\dagger$ creates an intracavity photon; $\hat{\sigma}^{\mu\nu}_p$ moves atom $p$ from state $\nu$ to state $\mu$; the allowed atomic states are $g$,$e$, and $f$; $U(r)$ is the atom-atom interaction potential, which exists only in the Rydberg state. Furthermore, $\omega_c=\nu_c+I\frac{\kappa}{2}$ with $\nu_c$ the cavity frequency and $\kappa$ the cavity linewidth. Similarly $\omega_e=\nu_e+I \frac{\Gamma}{2}$, $\omega_r=\nu_r+I\frac{\gamma_r}{2}$; note that, as written, $\nu_i$ is the angular frequency corresponding to the state with (complex) energy $\hbar \omega_i$, not the related cycle frequency.

\subsection{Resonator Transmission Spectrum}
In the linear regime, the resonator transmitted field amplitude is given by $t(\omega)\equiv\langle 0|\hat{a}|\psi_1\rangle$, where $|\psi_1\rangle=[\hat{H}-\hat{n}\omega]^{-1}\hat{V}|0\rangle$. We would like to avoid having to perform a matrix inversion or linear solve to compute $|\psi_1\rangle$; to this end we rewrite the previous equations (noting that within the single excitation manifold, $\hat{n}=\unit$): 
\begin{equation}
\label{eq:psi1def}
[\hat{H}-\omega\unit]|\psi_1\rangle=\Omega_p \hat{a}^\dagger|0\rangle
\end{equation}

Next we write $|\psi_1\rangle$ in a basis of single-excitation states:
\begin{equation}
\label{eq:psi1decomp}
|\psi_1\rangle=\left[A_c \hat{a}^\dagger+\sum_m\left( A_e^m \hat{\sigma}^{eg}_m+A_r^m \hat{\sigma}^{rg}_m\right)\right]|0\rangle
\end{equation}

We now plug Eqn. \ref{eq:psi1decomp} into Eqn. \ref{eq:psi1def}, yielding the following system of linear equations for $A_c$, $A_e^m$ and $A_r^m$:

\begin{eqnarray}
\Omega_p&=&A_c(\omega_c-\omega)+\sum_m {g_m^*A_e^m}\\
0       &=&A_e^m(\omega_e-\omega)+g_m A_c+\Omega_m^* A_r^m\nonumber\\
0       &=&A_r^m(\omega_r-\omega)+\Omega_m A_e^m\nonumber
\end{eqnarray}

We may now solve for $A_r^m$ in terms of $A_e^m$: $A_r^m=-A_e^m\frac{\Omega_m}{\omega_r-\omega}$, and plug into the equation for $A_e^m$, and solve in terms of $A_c$:
\begin{equation}
A_e^m=-A_c \frac{g_m}{(\omega_e-\omega)-\frac{|\Omega_m|^2}{\omega_r-\omega}}
\end{equation}

Finally we plug this equation into the equation determining $A_c$ in terms of the $A_e^m$, and solve, arriving at the following expression for the cavity transmission:
\begin{equation}
\label{eq:teqn}
\boxed{t(\omega)=A_c=\frac{\Omega_p}{(\omega_c-\omega)-\sum_m\frac{|g_m|^2}{(\omega_e-\omega)-\frac{|\Omega_m|^2}{\omega_r-\omega}}}}
\end{equation}

\subsection{Resonator $g_2$: Lowest Order non-Linear Transmission}
To compute the first nonlinear contribution to the transmission, we need $|\psi_2\rangle=[\hat{H}-\hat{n}\omega]^{-1}\hat{V}[\hat{H}-\hat{n}\omega]^{-1}\hat{V}|0\rangle=[\hat{H}-2\omega \unit]^{-1}\hat{V}|\psi_1\rangle$. Following the preceding section, we rewrite this as: $[\hat{H}-2\omega \unit]|\psi_2\rangle=\hat{V}|\psi_1\rangle$, and expand $|\psi_2\rangle$ in a two-excitation basis:

\begin{eqnarray}
\label{eq:psi2decomp}
|\psi_2\rangle=     & &B_{cc} \frac{1}{\sqrt{2}}\hat{a}^{\dagger2}|0\rangle\\
					&+&\sum_m{\left( B_{ce}^m \hat{a}^\dagger \hat{\sigma}^{eg}_m+B_{cr}^m \hat{a}^\dagger \hat{\sigma}^{rg}_m\right)}|0\rangle\nonumber\\
					&+&\sum_{n\neq m}{(B_{ee}^{mn}\hat{\sigma}^{eg}_m\hat{\sigma}^{eg}_n+B_{er}^{mn}\hat{\sigma}^{eg}_m\hat{\sigma}^{rg}_n+B_{rr}^{mn}\hat{\sigma}^{rg}_m\hat{\sigma}^{rg}_n)}|0\rangle\nonumber
\end{eqnarray}

We can now write out the equations of motion:
\begin{eqnarray}
\sqrt{2}\Omega_p A_c  &=&B_{cc}(2\omega_c-2\omega)+\sum_m {\sqrt{2}g_m^* B_{ce}^m}\\
        \Omega_p A_e^m&=&B_{ce}^m(\omega_c+\omega_e-2\omega)+\sqrt{2}g_m B_{cc}+\Omega_m^* B_{cr}^m+\sum_n{g_n^* B_{ee}^{mn}}\nonumber\\
        \Omega_p A_r^m&=&B_{cr}^m(\omega_c+\omega_r-2\omega)+\Omega_m B_{ce}^m+\sum_n {g_n^* B_{er}^{nm}}\nonumber\\
        0			  &=&B_{ee}^{mn}(2\omega_e-2\omega)+(g_n B_{ce}^m+g_m B_{ce}^n)+(\Omega_n^* B_{er}^{mn}+\Omega_m^* B_{er}^{nm})\nonumber\\
        0			  &=&B_{er}^{mn}(\omega_e+\omega_r-2\omega)+g_m B_{cr}^n+\Omega_n B_{ee}^{mn}+\Omega_m^* B_{rr}^{mn}\nonumber\\
        0			  &=&B_{rr}^{mn}(2\omega_r-2\omega+U_{mn})+\Omega_m B_{er}^{mn}+\Omega_n B_{er}^{nm}\nonumber
\end{eqnarray}

Where $U_{mn}\equiv U(|\vec{x}_m-\vec{x}_n|)$ is the interaction potential between Rydberg atoms $m$ and $n$. To proceed, we next rewrite $B_{er}^{mn}$ in terms of symmetrized coefficients $B_{erS}^{mn}=\frac{\Omega_m B_{er}^{mn}+\Omega_n B_{er}^{nm}}{\sqrt{|\Omega_m|^2+|\Omega_n|^2}}$, and anti-symmetrized coefficients $B_{erA}^{mn}=\frac{\Omega_n^* B_{er}^{mn}-\Omega_m^* B_{er}^{nm}}{\sqrt{|\Omega_m|^2+|\Omega_n|^2}}$, the latter of which are dark to $B_{rr}^{mn}$. The equations of motion become:
\begin{eqnarray}
\sqrt{2}\Omega_p A_c  &=&B_{cc}(2\omega_c-2\omega)+\sum_m {\sqrt{2}g_m^* B_{ce}^m}\\
        \Omega_p A_e^m&=&B_{ce}^m(\omega_c+\omega_e-2\omega)+\sqrt{2}g_m B_{cc}+\Omega_m^* B_{cr}^m+\sum_n{g_n^* B_{ee}^{mn}}\nonumber\\
        \Omega_p A_r^m&=&B_{cr}^m(\omega_c+\omega_r-2\omega)+\Omega_m B_{ce}^m+\sum_n {g_n^*(-\cos{\theta_{mn}} B_{erA}^{mn}+\sin{\theta_{mn}} B_{erS}^{mn})}\nonumber\\
        0			  &=&B_{ee}^{mn}(2\omega_e-2\omega)+(g_n B_{ce}^m+g_m B_{ce}^n)+\Omega_{mn}(\sin{2\theta_{mn}} B_{erS}^{mn}-\cos{2\theta_{mn}}B_{erA}^{mn})\nonumber\\
        0			  &=&B_{erS}^{mn}(\omega_e+\omega_r-2\omega)+g_m \cos{\theta_{mn}}B_{cr}^n+g_n \sin{\theta_{mn}}B_{cr}^m+\Omega_{mn} B_{rr}^{mn}+\Omega_{mn}\sin{2\theta_{mn}} B_{ee}^{mn} \nonumber\\
        0			  &=&B_{erA}^{mn}(\omega_e+\omega_r-2\omega)+g_m \sin{\theta_{mn}} B_{cr}^n-g_n \cos{\theta_{mn}} B_{cr}^m-\Omega_{mn}\cos{2\theta_{mn}}B_{ee}^{mn}\nonumber\\
        0			  &=&B_{rr}^{mn}(2\omega_r-2\omega+U_{mn})+\Omega_{mn}B_{erS}^{mn}\nonumber
\end{eqnarray}

Where $\cos{\theta_{mn}}\equiv\frac{\Omega_m}{\Omega_{mn}}$, $\sin{\theta_{mn}}\equiv\frac{\Omega_n}{\Omega_{mn}}$, $\Omega_{mn}\equiv\sqrt{\Omega_m^2+\Omega_n^2}$ and we assume $\Omega_m$,$\Omega_n$ real. We can now follow the procedure that we employed in the preceding section and explicitly eliminate $B_{rr}^{mn}$ in terms of $B_{erS}^{mn}$, and then $B_{erS}^{mn}$ and $B_{erA}^{mn}$, and finally $B_{ee}^{mn}$:

\begin{eqnarray}
B_{rr}^{mn} &=&-B_{erS}^{mn} \frac{\Omega_{mn}}{2\omega_r-2\omega+U_{mn}}\\
B_{erA}^{mn}&=&-\frac{g_m\sin{\theta_{mn}} B_{cr}^n-g_n\cos{\theta_{mn}} B_{cr}^m-\Omega_{mn}\cos{2\theta_{mn}}B_{ee}^{mn}}{\omega_e+\omega_r-2\omega}\nonumber\\
B_{erS}^{mn}&=&-\frac{g_m\cos{\theta_{mn}} B_{cr}^n+g_n\sin{\theta_{mn}} B_{cr}^m+\Omega_{mn}\sin{2\theta_{mn}}B_{ee}^{mn}}{\omega_e+\omega_r-2\omega-\frac{\Omega_{mn}^2}{2\omega_r-2\omega+U_{mn}}}\nonumber\\
B_{ee}^{mn} &=&\frac{(g_n B_{ce}^m+g_m B_{ce}^n)-\Omega_{mn}(\sin{2\theta_{mn}} \frac{g_m\cos{\theta_{mn}} B_{cr}^n+g_n\sin{\theta_{mn}} B_{cr}^m}{\omega_e+\omega_r-2\omega-\frac{\Omega_{mn}^2}{2\omega_r-2\omega+U_{mn}}}+\cos{2\theta_{mn}}\frac{g_m\sin{\theta_{mn}} B_{cr}^n-g_n\cos{\theta_{mn}} B_{cr}^m}{\omega_e+\omega_r-2\omega})}{2\omega_e-2\omega-\frac{\Omega_{mn}^2 \sin^2{2\theta_{mn}}}{\omega_e+\omega_r-2\omega-\frac{\Omega_{mn}^2}{2\omega_r-2\omega+U_{mn}}}-\frac{\Omega_{mn}^2 \cos^2{2\theta_{mn}}}{\omega_e+\omega_r-2\omega}}\nonumber
\end{eqnarray}

Plugging these equations into the remaining equations determining $B_{cc}$,$B_{ce}^m$, and $B_{cr}^m$ explicitly eliminates the $\sim 2N^2$ equations where neither of two excitations is a cavity photon (both are atomic excitations), leaving only the $2N+1$ equations where either both excitations are resonator photons (one equation) or one is a resonator photon and the other an atomic excitation. Note that these new equations are completely dense (not sparse), as $B_{ee}^{mn}$ directly couples $B_{cr}^m$ to $B_{cr}^n$ and $B_{ce}^m$ to $B_{ce}^n$.

In the simple case that the control beam is uniform across the sample, $\Omega_m=\Omega$ independent of $m$, then we have $\theta_{mn}=\frac{\pi}{4}$, and the equations simplify substantially:
\begin{eqnarray}
B_{rr}^{mn} &=&-B_{erS}^{mn} \frac{\sqrt{2}\Omega}{2\omega_r-2\omega+U_{mn}}\\
B_{erA}^{mn}&=&-\frac{\frac{1}{\sqrt{2}}(g_m B_{cr}^n-g_n B_{cr}^m)}{\omega_e+\omega_r-2\omega}\nonumber\\
B_{erS}^{mn}&=&-\frac{\frac{1}{\sqrt{2}}(g_m B_{cr}^n+g_n B_{cr}^m)+\sqrt{2}\Omega B_{ee}^{mn}}{\omega_e+\omega_r-2\omega-\frac{2\Omega^2}{2\omega_r-2\omega+U_{mn}}}\nonumber\\
B_{ee}^{mn} &=&\frac{(g_n B_{ce}^m+g_m B_{ce}^n)-\Omega\left(\frac{g_m B_{cr}^n+g_n B_{cr}^m}{\omega_e+\omega_r-2\omega-\frac{2\Omega^2}{2\omega_r-2\omega+U_{mn}}}\right)}{2\omega_e-2\omega-\frac{2\Omega^2}{\omega_e+\omega_r-2\omega-\frac{2\Omega^2}{2\omega_r-2\omega+U_{mn}}}}\nonumber
\end{eqnarray}

We can now plug these simplified equations into the remaining equations, to arrive $2N+1$ coupled equations which must be solved numerically (where we assume the $g_n$ are real):

\begin{eqnarray}
\sqrt{2}\Omega_p A_c &=& (2\omega_c-2\omega) B_{cc}+\sqrt{2}\sum_m g_m B_{ce}^m\\
\Omega_p A_e^m       &=& (\omega_c+\omega_e+\sum_n{\frac{|g_n|^2}{Z_{mn}}}-2\omega) B_{ce}^m+\sqrt{2}g_m B_{cc}+\Omega B_{cr}^m\left(1+\sum_n{\frac{|g_n|^2}{Z_{mn}Y_{mn}}}\right)\nonumber\\
					 &+&\sum_n{\frac{g_m g_n}{Z_{mn}}\left(B_{ce}^m+B_{cr}^m\frac{\Omega}{Y_{mn}}\right)}\nonumber\\
\Omega_p A_r^m		 &=& (\omega_c+\omega_r+\sum_n{|g_n|^2\left(\frac{\Omega^2}{Z_{mn} Y_{mn}^2}+\frac{1}{2Y_{mn}}-\frac{1}{2(\omega_e+\omega_r-2\omega)}\right)}-2\omega)B_{cr}^m\nonumber\\
					 &+&\Omega B_{ce}^m\left(1+\sum_n\frac{g_n^2}{Z_{mn}Y_{mn}} \right)+\sum_n{g_m g_n\left(\frac{\Omega^2}{Z_{mn}Y_{mn}^2}+\frac{1}{2Y_{mn}}+\frac{1}{2(\omega_r+\omega_e-2\omega)}\right)B_{cr}^n}\nonumber\\
					 &+&\Omega\sum_m{B_{ce}^n\frac{g_m g_n}{Z_{mn}Y_{mn}}}\nonumber
\end{eqnarray}

Where $Y_{mn}\equiv 2\omega-(\omega_e+\omega_r)-\frac{2\Omega^2}{2\omega-(2\omega_r+U_{mn})}$, and $Z_{mn}\equiv 2(\omega-\omega_e-\frac{2\Omega^2}{Y_{mn}})$.

It is this last set of equations that we solve numerically, for many ensembles of randomly sampled atoms over a cross-section larger than the resonator mode waist, to generate the ``model'' curves for the $g_2$.
\end{widetext}
\end{document}